\documentclass[conference]{IEEEtran}
\usepackage[utf8]{inputenc}
\usepackage[english]{babel}
\usepackage{amsmath,amsfonts}
\usepackage[pscoord]{eso-pic}
\PassOptionsToPackage{detect-all, per-mode=symbol, range-units=single, range-phrase=~to~}{siunitx}
\usepackage{siunitx}
\usepackage{nicefrac}
\usepackage{orcidlink}
\usepackage[noabbrev,capitalise]{cleveref}
\usepackage{float}
\usepackage[caption=false]{subfig}
\usepackage{lipsum}
\usepackage{placeins}
\usepackage{xfrac}
\usepackage{pifont}
\usepackage{threeparttable}
\usepackage{booktabs}
\usepackage{colortbl}
\usepackage{tabularx}
\usepackage{makecell}
\usepackage{graphicx}
\usepackage{xspace}
\usepackage{algorithmic}
\usepackage{graphicx}
\usepackage{textcomp}
\usepackage{xcolor}
\usepackage{array}
\usepackage{setspace}
\usepackage{enumitem}
\usepackage{url}
\usepackage[font=footnotesize,labelfont=bf]{caption}
\usepackage{gensymb}
\usepackage{rotating}
\usepackage{tabularray}
\usepackage{rotating}
\usepackage[acronym]{glossaries}

\definecolor{light-gray}{gray}{0.75}

\newcommand{\reb}[1]{{\color{black} #1}}

\newcommand{\x}{$\times$}

\renewcommand{\subsubsection}[1]{\paragraph*{\textbf{#1}}}

\DeclareSIUnit{\x}{\!\ensuremath{\times}}
\DeclareSIUnit\bit{b}
\DeclareSIUnit\GE{GE}
\DeclareSIUnit\kGE{\kilo\GE}
\DeclareSIUnit\MGE{\mega\GE}
\def\reviewpass{v0.3.0}
\begin{document}

%%%%%%%%%%%%%%%%%%%%%%
%%   VERSIONING     %%
%%%%%%%%%%%%%%%%%%%%%%

% version
\ifx\showrevision\undefined
    \newcommand{\todo}[1]{{#1}}
    \newcommand{\ph}[1]{{#1}}
\else
    \newcommand{\todo}[1]{{\textcolor{ieee-bright-red-80}{#1}}\PackageWarning{TODO:}{#1!}}
    \newcommand{\ph}[1]{{\textcolor{light-gray}{#1}}\PackageWarning{PH:}{#1!}}
    \AddToShipoutPictureFG{%
        \put(%
            8mm,%
            \paperheight-1.5cm%
            ){\vtop{{\null}\makebox[0pt][c]{%
                \rotatebox[origin=c]{90}{%
                    \huge\textcolor{ieee-bright-red-80!75}{\reviewpass}%
                }%
            }}%
        }%
    }
    \AddToShipoutPictureFG{%
        \put(%
            \paperwidth-6mm,%
            \paperheight-1.5cm%
            ){\vtop{{\null}\makebox[0pt][c]{%
                \rotatebox[origin=c]{90}{%
                    \huge\textcolor{ieee-bright-red-80!30}{Unpublished - Confidential - Draft - Copyright 2023}%
                }%
            }}%
        }%
    }
\fi

% Acronyms
\newacronym{dtm}{DTM}{Dynamic Thermal Management}
\newacronym{dpm}{DPM}{Dynamic Power Management}
\newacronym{dptm}{DPTM}{Dynamic Power and Thermal Management}
\newacronym{stm}{STM}{Static Thermal Management}
\newacronym{hw}{HW}{hardware}
\newacronym{sw}{SW}{software}
\newacronym{ca}{CA}{command/address}
\newacronym{ip}{IP}{intellectual property}
\newacronym{ddr}{DDR}{double data rate}
\newacronym{lpddr}{LPDDR}{low-power DDR}
\newacronym{rpc}{RPC}{reduced pin count}
\newacronym{dma}{DMA}{direct memory access}
\newacronym{axi}{AXI}{Advanced eXtensible Interface}
\newacronym{dram}{DRAM}{Dynamic Random Access Memory}
\newacronym[firstplural=static random access memories (SRAMs)]{sram}{SRAM}{Static Random Access Memory}
\newacronym{edram}{eDRAM}{embedded DRAM}
\newacronym[firstplural=systems on chip (SoCs)]{soc}{SoC}{System on Chip}
\newacronym{mpsoc}{MPSoC}{Multi-Processor System on Chip}
\newacronym{hesoc}{HeSoC}{Heterogeneous System on Chip}
\newacronym{sip}{SiP}{System in Package}
\newacronym{fpga}{FPGA}{Field-Programmable Gate Array}
\newacronym{asic}{ASIC}{Application-Specific Integrated Circuit}
\newacronym{phy}{PHY}{physical layer}
\newacronym{ml}{ML}{Machine Learning}
\newacronym{iot}{IoT}{Internet of Things}
\newacronym{foss}{FOSS}{free and open source}
\newacronym{cmos}{CMOS}{Complementary Metal-Oxide-Semiconductor}
\newacronym{sut}{SUT}{system under test}
\newacronym{isut}{ISUT}{integrated system under test}
\newacronym{rtl}{RTL}{register transfer level}
\newacronym{hil}{HIL}{Hardware in the Loop}
\newacronym{pil}{PIL}{Processor in the Loop}
\newacronym{fil}{FIL}{FPGA in the loop}
\newacronym{mil}{MIL}{Model in the Loop}
\newacronym{sil}{SIL}{Software in the Loop}
\newacronym{hpc}{HPC}{High Performance Computing}
\newacronym{mcu}{MCU}{Microcontroller Unit}
\newacronym{fub}{FUB}{Functional Unit Block}
\newacronym{ecu}{ECU}{Electronic Control Unit}
\newacronym{dcu}{DCU}{Domain Control Unit}
\newacronym{adas}{ADAS}{advanced driver-assistance system}
\newacronym{fame}{FAME}{FPGA Architecture Model Execution}
\newacronym{pl}{PL}{Programmable Logic}
\newacronym{ps}{PS}{Processing System}
\newacronym{apu}{APU}{Application Processing Unit}
\newacronym{ocm}{OCM}{On-Chip Memory}
\newacronym{pcs}{PCS}{Power Controller System}
\newacronym{pcf}{PCF}{Power Control Firmware}
\newacronym{bram}{BRAM}{block RAM}
\newacronym{lut}{LUT}{look-up table}
\newacronym{ff}{FF}{flip-flop}
\newacronym{fsbl}{FSBL}{First Stage BootLoader}
\newacronym{pvt}{PVT}{Process, Voltage, Temperature}
\newacronym{hls}{HLS}{high-level synthesis}
\newacronym{mqtt}{MQTT}{Message Queuing Telemetry Transport}
\newacronym{cots}{COTS}{commercial off-the-shelf}
\newacronym{cpu}{CPU}{Central Processing Unit}
\newacronym{gpu}{GPU}{Graphic Processing Unit}
\newacronym{ibmocc}{IBM OCC}{IBM On-Chip Controller}
\newacronym{clic}{CLIC}{Core-Local Interrupt Controller}
\newacronym{clint}{CLINT}{Core-Local Interruptor}
\newacronym{scmi}{SCMI}{System Control and Management Interface}
\newacronym{os}{OS}{Operating System}
\newacronym{ospm}{OSPM}{Operating System Power Management}
\newacronym{mimo}{MIMO}{Multiple-Input Multiple-Output}
\newacronym{siso}{SISO}{Single-Input Single-Output}
\newacronym{bmc}{BMC}{Board Management Controller}
\newacronym{qos}{QoS}{Quality of Service}
\newacronym{tdp}{TPD}{Thermal Design Power}
\newacronym{dvfs}{DVFS}{Dynamic Voltage and Frequency Scaling}
\newacronym{dfs}{DFS}{Dynamic Frequency Scaling}
\newacronym{dvs}{DVS}{Dynamic Voltage Scaling}
\newacronym{rtu}{RTU}{Real Time Unit}
\newacronym{pe}{PE}{Processing Element}
\newacronym{noc}{NoC}{Network on Chips}
\newacronym{pid}{PID}{Proportional Integral Derivative}
\newacronym{sota}{SotA}{State-of-the-Art}
\newacronym{fpu}{FPU}{Floating Point Unit}
\newacronym{pcu}{PCU}{Power Control Unit}
\newacronym{scp}{SCP}{System Control Processor}
\newacronym{mcp}{MCP}{Manageability Control Processor}
\newacronym{occ}{OCC}{On-Chip Controller}
\newacronym{smu}{SMU}{System Management Unit}
\newacronym{ap}{AP}{Application-class Processors}
\newacronym{vrm}{VRM}{Voltage Regulator Module}
\newacronym{pfct}{PFCT}{Periodic Frequency Control Task}
\newacronym{pvct}{PVCT}{Periodic Voltage Control Task}
\newacronym{ipc}{IPC}{Instructions per Cycle}
\newacronym{simd}{SIMD}{Single Instruction, Multiple Data}
\newacronym{mctp}{MCTP}{Management Component Transport Protocol}
\newacronym{pldm}{PLDM}{Platform Level Data Model}
\newacronym{rtos}{RTOS}{Real-Time OS}
\newacronym{hlc}{HLC}{High-Level Controller}
\newacronym{llc}{LLC}{Low-Level Controller}
\newacronym{acpi}{ACPI}{Advanced Configuration and Power Interface}
\newacronym{pdn}{PDN}{Power Delivery Network}
\newacronym{ewma}{EWMA}{Exponential Weight Moving Average}
\newacronym{ppa}{PPA}{power, performance and area}
\newacronym{pcb}{PCB}{Printed Circuit Board}
%\newacronym{soa}{SoA}State of the Art}
%
\newacronym{dsa}{DSA}{Domain-Specific Architecture}
\newacronym{ha}{HA}{Hardware Accelerator}
\newacronym{pmca}{PMCA}{Programmable Multi-Core Accelerator}

\newacronym[longplural={high-bandwidth memories}]{hbm}{HBM}{high-bandwidth memory}
\newacronym{rapl}{RAPL}{Running Average Power Limit}
\newacronym{tsv}{TSV}{through silicon via}
\newacronym{fet}{FET}{Field Effect Transistor}

\newacronym{fll}{FLL}{Frequency Locked Loop}
\newacronym{pll}{PLL}{Phase Locked Loop}

\newacronym{oca}{EBA}{Enhanced Baseline Algorithm}
\newacronym{fca}{FCA}{Fuzzy-ispired Iterative Control Algorithm}
\newacronym{vba}{VBA}{Voting Box Algorithm}
\newacronym{ba}{BA}{Baseline Algorithm}

\newacronym{dt}{DT}{Dispatch Throttling}

\newacronym{ldo}{LDO}{Low-Dropout Regulator}

\newacronym{tim}{TIM}{Thermal Interface Material}
\newacronym{mpc}{MPC}{Model Predictive Control}

%%%%%%%%%%%%%%%%%%%%%%
%%   FRONT MATTER   %%
%%%%%%%%%%%%%%%%%%%%%%

\title{Modeling and Controlling Many-Core HPC Processors: an Alternative to PID and Moving Average Algorithms
} 

%TODO: complete with correct orcid links
\ifx\blind\undefined


\author{
    \IEEEauthorblockN{%
    Giovanni Bambini\textsuperscript{\textasteriskcentered}, % 
    Alessandro Ottaviano\textsuperscript{\textdagger}, % 
    Christian Conficoni\textsuperscript{\textasteriskcentered}, % 
    Andrea Tilli\textsuperscript{\textasteriskcentered}, % 
    Luca Benini\textsuperscript{\textasteriskcentered}\textsuperscript{\textdagger}, \\% 
    Andrea Bartolini\textsuperscript{\textasteriskcentered}, % 
    }
   \IEEEauthorblockA{
        \textasteriskcentered~\textit{Department of Electrical, Electronic, and Information Engineering, University of Bologna}, Italy \\
       \textdagger~\textit{Integrated Systems Laboratory, ETH Zurich}, Switzerland \\
        \
    }
}

\maketitle

\begin{abstract}

The race towards performance increase and computing power has led to chips with heterogeneous and complex designs, integrating an ever-growing number of cores on the same monolithic chip or chiplet silicon die. %
Higher integration density, compounded with the slowdown of technology-driven power reduction, implies that power and thermal management become increasingly relevant. %
Unfortunately, existing research lacks a detailed analysis and modeling of thermal, power, and electrical coupling effects and how they have to be jointly considered to perform dynamic control of complex and heterogeneous \glspl{mpsoc}.
To close the gap, in this work, we first provide a detailed thermal and power model targeting a modern \gls{hpc} \gls{mpsoc}. We consider real-world coupling effects such as actuators' non-idealities and the exponential relation between the dissipated power, the temperature state, and the voltage level in a single processing element. We analyze how these factors affect the control algorithm behavior and the type of challenges that they pose. 
Based on the analysis, we propose a thermal capping strategy inspired by Fuzzy control theory to replace the state-of-the-art PID controller, as well as a root-finding iterative method to optimally choose the shared voltage value among cores grouped in the same voltage domain.
We evaluate the proposed controller with model-in-the-loop and hardware-in-the-loop co-simulations. 
We show \reb{an improvement over state-of-the-art methods of up to $5\times$ the maximum exceeded temperature while providing an average of $3.56\%$} faster application execution runtime across all the evaluation scenarios.

\end{abstract}

%\keywords{Modeling, Control, Nonlinear systems}

%%%%%%%%%%%%%%%%%%
%%   CONTENT   %%%
%%%%%%%%%%%%%%%%%%

\section{Introduction} \label{sec:intro}

%%%%%%%%%%%%%%%%%%%%%%%%%%%%%
%%% Brief Introduction on HPC
Modern high-performance \glspl{cpu} are heterogeneous architectures that feature many processing elements integrated on a monolithic silicon die or silicon chiplet. %
%Processing elements are either traditional general-purpose machines or \glspl{dsa} such as general-purpose \glspl{pmca}, dedicated \glspl{ha} engines, and \glspl{gpu}.
These complex and heterogeneous architectures require a runtime dynamic thermal and power management strategy that allows fine-grained monitoring and control of the application workloads to either keep the system within well-specified \gls{tdp} and minimize power spikes and temperature hot spots.
Growing attention is indeed being paid to energy-efficient large-scale \glspl{hpc}.

%%%%%%%%%%%%%%%%%%%%%%%%%%%%%
%%% Introduce the Challenges of HPC Thermal and Power Control

The control problem introduced by modern many-core architecture is challenging: it is a non-linear problem with fast dynamics, involving multiple coupled control elements that are subjected to variable constraints within a system with high-amplitude and high-frequency disturbances (\cref{ssec:ctrl_challenges}). These constraints have to be enforced while aiming at selecting the operating point at the highest power-affordable performance for the system executing an application workload. 
Furthermore, controlled inputs may be coupled due to the many-cores design that pushes the limit of hardware integration for voltage regulators, forcing several cores to share a single voltage through the same power domain rail.

%%%%%%%%%%%%%%%%%%%%%%%%%%%%%
%%% Introduce the control Flow and the difference between HLC and LLC

Traditional dynamic power management strategies range from architectural techniques involving gate-level clock/power gating to \emph{runtime active control} carried out by higher layers of the \gls{os} and firmware software stacks~\cite{Tilli2022}.
\reb{In the latter scenario, which is the focus of this work, the control signal travels through several control blocks in cascade steps, as shown in \cref{fig:cascade}. Each control block is characterized by a certain level of system abstraction and knowledge of the surroundings. In the current literature (\cref{sec:related_works}), we identify three main categories of controllers and algorithms according to their execution context: a class of controllers that oversee multiple systems (e.g. whole data centers, or entire nodes),} a class where the control action originates within a single system through high-level controllers, such as the \gls{ospm} running on the application-class cores or on the off-chip \gls{bmc}, and a class where the control action originates from a dedicated unit integrated within the \glspl{mpsoc}. % and tasked to relieve the power and thermal management burden from the \glspl{hlc}. 
The latter are typically low-power embedded microcontrollers in the \gls{mpsoc}'s \textit{uncore} domain, with privileged access and high bandwidth control over the system architecture. Hence, they are referred to as \gls{llc} in this work~\cite{PERCOREPSTATE, PWRMM_TECHNIQUES}, while the other two categories are referred to as global and local \glspl{hlc}.

\begin{figure*}[t]
	\centering	\includegraphics[width=0.9\linewidth]{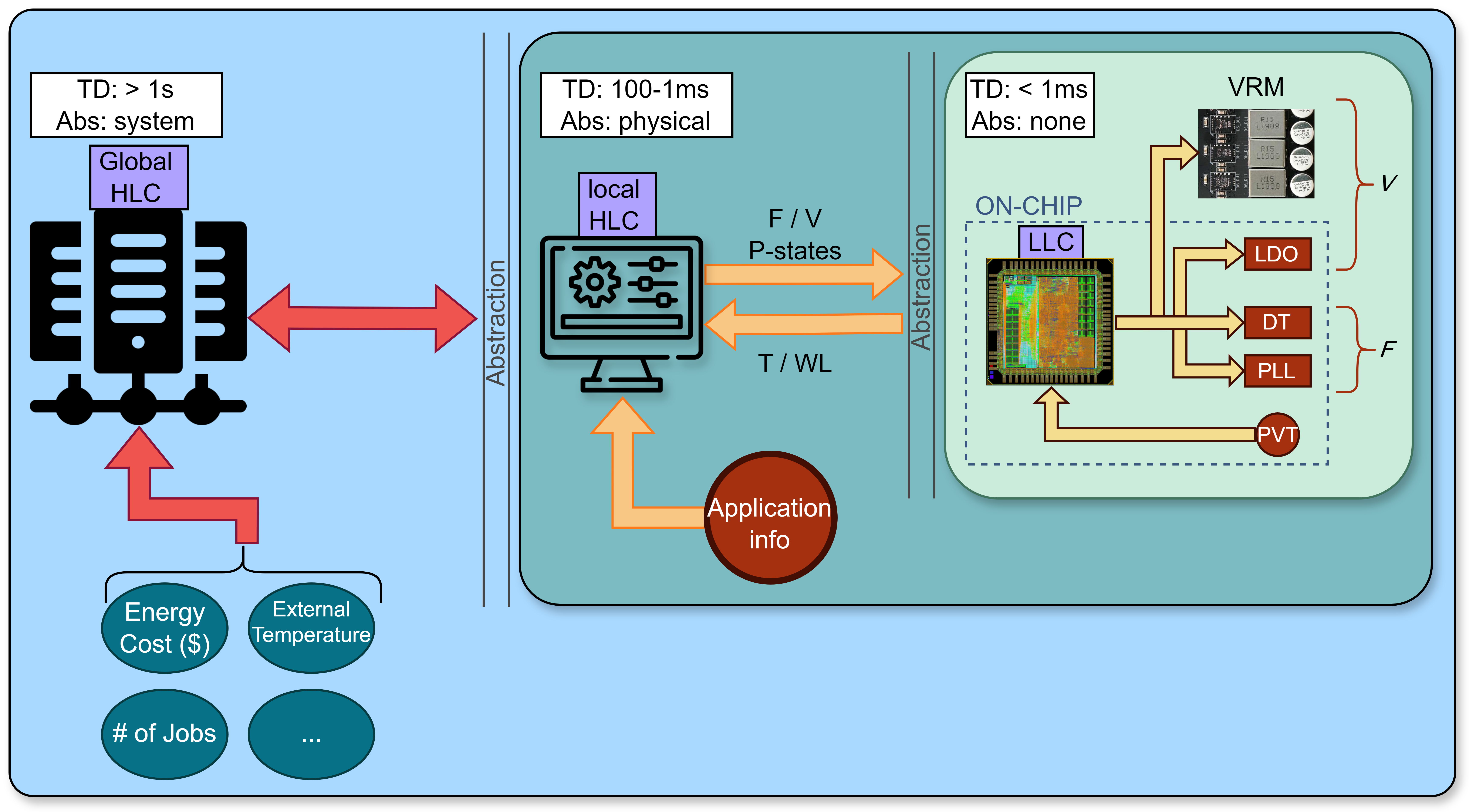}
	\caption{\label{fig:cascade} \reb{Architecture of the cascade control blocks in an \gls{hpc} system. The three blocks have separate time domains, their own abstraction of the system to be controlled, and distinct external information depending on their scope. In particular, the \gls{llc} (inner light green block) controls the low-level,  physical parts of the system (\acrfull{vrm}, \acrfull{ldo}, \acrfull{dt}, and \acrfull{pll}) based on the target operating points given by the local \gls{hlc}. The \gls{llc} also fetches information from the \acrfull{pvt} and other sensors and communicates it to the \gls{hlc}. The \gls{hlc} (middle dark green block) takes information from the \gls{llc} about the system as well as information from the executing application and computes a set of operating points. Finally, the global \gls{hlc} (external light blue block) takes information from the system and from the surroundings and communicates it to a performance/energy efficiency target to each local \gls{hlc}.}}
	%\vspace{-15pt}
\end{figure*}

%evidenziare che ci mettiamo / testiamo sull'hardware (fpga) HIL
%evitiamo di dire nel cluster o sul manager core -> non importa dove viene eseguito ai fini del controllo
%dire che siamo i primi a solve il problema del coupling e in piu su un embedded controller

\reb{Most of the \gls{sota} focuses on \glspl{hlc}~\cite{TPDS, HPC-survey19}. We conjecture this happens because of the lack of an open-source \gls{hw}/\gls{sw} research framework able to capture the complexity and interfaces of an \gls{llc} subsystem integrated into a modern \gls{hpc} \gls{mpsoc}.%, and the realization by the researchers, that otherwise their work would be never implemented in an industry chip.
%This is about to change with the rise of open-source ISA \ref{TODO} and general-purpose chips \ref{TODO}, which are currently lacking \gls{sota} research on the \gls{llc}, and for this reason, we are providing a detailed analysis on the creation of such framework (as well as providing the one we used in this work)
}.

\glspl{llc} have also become increasingly important in the context of advanced power management. It has become clear the trend, originating from industry, of abstracting power and system management tasks away from the \glspl{hlc} \glspl{ap} towards the \glspl{llc}~\cite{scp_repo, schone_energy_2019} intended as control systems that are closer to the controlled hardware components, and increasingly performant \glspl{llc}, such as parallel microcontrollers, are able to address multi-variable optimization scenarios without the need to reduce the complexity of the control problem, simultaneously retaining low-power, and energy-efficiency and real-time bounds required by the \textit{uncore} domain and the nature of the control problem~\cite{UNCORE}.
\reb{The objective of this work is to provide an overview and analysis of the \gls{llc} control problem and its challenges, as well as the design of an open-source algorithm developed in a real-world setting to serve as proxy for future research.}

%%%%%%%%%%%%%%%%%%%%%%%%%%%%%
%%% Wrap up

%%%%%%%%%%%%%%%%%%%%%%%%%%%
%%% Contributions:

To the best of the authors' knowledge, little research has addressed the control problem of \gls{hpc} systems with full coupling between power, frequency, voltage, and temperature. As shown in \cref{sec:related_works}, \gls{sota} controllers from industry and academia employ simplified control policies often unable to address the challenges of novel heterogeneous many-core architectures. %

In this manuscript, we significantly extend the conference work in Bambini et al.~\cite{CCTA_bambini} and propose the first analysis and implementation of a cascade reactive control policy that tackles the whole multi-variable control problem with actuators coupling and non-idealities. 
We provide a cost assessment based on \gls{mil} and \gls{hil} emulations. In the latter, we implement the algorithm on a \gls{fpga}-\gls{soc} framework codesigned with an open-source, RISC-V embedded \gls{llc}~\cite{Ottaviano2023ControlPULPAR} to account for the physical state time scale of the control variables.
This work provides the following contributions:

\begin{itemize}
    \item We comprehensively describe a multi-variable power and thermal control problem in a modern \gls{hpc} \gls{mpsoc} (\cref{sec:ctrlobj} and \cref{sec:methods}). The control targets a \gls{llc} that reacts to high-level decisions from the surrounding \glspl{hlc} (\gls{ospm}, \gls{bmc}) and dispatches \gls{dvfs} operating points while a workload is being executed on the controlled system. We present how to model the computational part of that system considering all the couplings occurring within the same component and between neighboring components. 
    %We show how the problem translates into a system of partial differential equations that we aim to solve with root-finding iterative methods (Sec. \ref{sec:method}).
    \item We extend the control algorithm presented in~\cite{ICCD_bambini, CCTA_bambini} to account for variable coupling and introduce a thermal capping approach inspired by the Fuzzy control theory. Finally, we address the shared voltage coupling with an iterative, root-finding method. 
    The resulting improved reactive control policy outperforms the \gls{sota} reactive PID under multiple different scenarios (\cref{sec:results}) in a \gls{mil}-based emulation framework.
    %\item We analyze the challenges from a control point of view laid in the relation between F and V in modern \gls{hpc} \gls{mpsoc}s. We present a new approach to tackle these challenges with a root-finding iterative method (Sec. \ref{sec:method})
    \item We evaluate the feasibility of the designed algorithms on a RISC-V open-source \gls{llc} controller (\cref{sec:results}). The \gls{hw} controller is mapped on a modern \gls{hesoc} \gls{fpga} and relies on a \gls{hil} evaluation framework that includes the controller and a power/thermal model of the controlled plant.
\end{itemize}

\reb{This work is freely available at the following repositories:
\begin{itemize}
    \item \url{https://github.com/Ev3nt1ne/AechPeSi_lab}: the Matlab simulation environment and control part;
    \item \url{https://github.com/pulp-platform/control-pulp} and \url{https://github.com/pulp-platform/control_pulp_pcf}: the proposed parallel \gls{llc} controller, including open-source hardware, firmware, and control algorithm;
    \item \url{https://github.com/pulp-platform/pulp_hpc_cosim}: the simulation environment written in C for FPGA-based co-simulation.
\end{itemize}
}
\section{Related Works}\label{sec:related_works}

%scrivere i related work per i controllisti!
%dai related works devono uscire delle classi di controlli: es. ci sono papers che considerano questo legame? ed eseguite a sistema operativo e non compatibili con le capacità di calcolo del microcontrollore

%here adding citations for HLC and LLC according to the paper in the bibliography to not overload it.
%HLC: Zhang2, CESARINI, barto_main, Conf_22
%LLC: SANDYBRIDGE, ICCD, schone

The \reb{joint} control objective of \gls{hlc} and \gls{llc} controllers \reb{is to} perform power and thermal management on the system while maintaining a dynamic trade-off between maximum performance and energy efficiency~\cite{SANDYBRIDGE}. %
Nevertheless, they differ in their computing capabilities, safety requirements, and the accessibility to sensors and actuators surrounding the physical system under control~\cite{Zhang2}.

The \gls{hlc} is generally part of the \gls{os} routine, or a user-level application executing concurrently with the main system workload~\cite{taxonomy2011}. 
Leveraging directly the application-class \glspl{pe}, the \gls{hlc} has access to higher computing capabilities than the \gls{llc}~\cite{Ottaviano2023ControlPULPAR}, albeit limited by the contention with the executing workload. Furthermore, lower processor privilege levels (e.g., supervisor in the RISC-V community~\cite{RISCVSPEC2}) have typically restricted access to the system itself for security reasons. 
%Unlike the \gls{llc} that has direct access to the system, the \gls{hlc} accesses already processed sensor data from the system with a lower frequency~\cite{something?}. 
%The sensor data may vary asynchronously, and it may not be entirely accessible at once~\cite{Khan}. Also, the controllability of the actuators is limited relative to the control interface made available by the system manufacturer, while conflicting control inputs between \gls{hlc} and \gls{llc} may be resolved by choosing the latter or with \textcolor{red}{undefined} behavior.

For these reasons, \glspl{hlc} generally support complex control algorithms aimed at \textit{prediction} and \textit{optimality}, which take into account a comprehensive characterization of the surroundings of the controlled system, such as coupling with neighboring systems, the ambient temperature, and other external factors~\cite{HPC-survey19}, instead of focusing on specific aspects of the system itself. Their objective is shifted more towards power efficiency and optimal application performance trade-off than power and thermal capping. To this end, \glspl{hlc} use an abstraction of the physical system through common protocols such as \gls{acpi}~\cite{schone_energy_2019}.

\reb{\Gls{mpc} has been used to control the temperature, by computing an optimal trajectory based on well-crafted energy cost functions~\cite{TPDS, Tilli2022, gpgpu}. Model identification techniques~\cite{TCAS,Beneventi} including ones based on \gls{ml}~\cite{paperA,energy-aware} are largely used to develop complex and accurate \gls{hlc} controllers that are updated in real-time. A wide literature exists for \gls{ml}-based \gls{hlc} controllers which not only optimize energy efficiency and execution performance but also task allocation and threads management~\cite{ml-survey}.

A trend toward lightweight \gls{hlc} controllers exists, but it is mainly related to mobile platforms which differ significantly from \gls{hpc} systems by number of \glspl{pe} and power scales. In Bhat et al.~\cite{paperB}, a lightweight \gls{dptm} based on an offline system model identification is deployed on a mobile platform to improve performance and energy efficiency by leveraging the sensitivity of the workload to frequency changes. Fuzzy controllers have also been successfully used in this context~\cite{hessle-free, fuzzy-variation}.
}

While advanced research in this field focuses on \gls{hlc}, they traditionally rely on \glspl{llc} to apply their policies and to carry out fine-grain thermal and power capping \cite{arm_pwmgt}. %
As a matter of fact, end-to-end analysis and evaluations of \glspl{llc} are scarce in \gls{sota} literature due mostly to the work being done behind industry secret, and the lack of freely accessible \gls{llc} hardware, software, and emulation frameworks. 
\Glspl{llc} are indeed tailored to the system they are integrated into --- hence, more difficult to generalize --- but control abstraction and analysis are possible. \reb{This work focuses only on the analysis of \glspl{llc} in the \gls{hpc} application domain.}%

To the best of the authors' knowledge, there are four major \gls{hpc} actors in the industry market employing their own \gls{llc}: Intel, AMD, IBM, and ARM. %

Intel uses individual functional blocks' power consumption as the controlled variable and performs dynamic budget allocation to the various components~\cite{INTEL_COREM_PWR, INTEL_XEON5_PWR}. The power is computed through a model whose dependencies include leakage information, voltage, temperature, and an estimated workload.
The control is characterized by an \gls{ewma} algorithm over multiple time windows of the order of seconds. According to~\cite{INTEL_pguide}, other thermal control features act in parallel with the power capping.
One of Intel's flagship management techniques is \gls{rapl}, a hardware feature for fine-grained energy consumption monitoring and power/thermal capping with support for several power domains in the system, covering \glspl{cpu}, \glspl{gpu}, and \gls{dram} controllers. Monitoring happens through model-specific registers (MSRs), simple counters integrated within the architecture. %

According to~\cite{AMD, AMD_BKDG}, AMD relies on a distributed \gls{pid} control structure. Both the thermal and the power control are performed independently using the frequency as the controlled variable, while a voting-box mechanism is tasked to resolve the coupling among components. Given the similarity to Intel architectures, AMD's power and thermal management strategy resembles \gls{rapl} from an implementation point of view. %

IBM releases its Power9 OpenPOWER firmware as open-source~\cite{ibm_repo}. From the source code, a voting-box mechanism selects the most constraining frequency for the system, similar to the AMD strategy. The temperature and power consumption are controlled independently, with the frequency being the controlled variable.
Differently from AMD, IBM's control firmware has a centralized design, i.e. it gathers data to a central unit and performs one single algorithm computation using moving average filters and \glspl{pid}. One of the blocks entering the voting-box mechanism is based on workload estimation. %

Likewise, Arm releases the \gls{scp} firmware as open-source, which consists of a cascade design with weight-based power distribution. The thermal dispatching layer precedes the power reduction one. They use power as the controlled variable, a PID for thermal capping, and a two-stage power distribution with a power accumulation variable to distribute unused power to the core.

%Arm implements two independent \glspl{pcs} based on the Arm Cortex-M7 microcontroller, \gls{scp} and \gls{mcp}. The \gls{scp} provides power management functionality, while the \gls{mcp} supports communications functionality. 
%In Arm-based \glspl{soc} the interaction with the \gls{os} is handled by the \gls{scmi} protocol~\cite{SCMI}. \gls{scmi} provides a set of \gls{os}-agnostic standard \gls{sw} and \gls{hw} interfaces for power domain, voltage, clock, and sensor management through a shared, interrupt-driven mailbox system with the \gls{pcs}.

%\textcolor{red}{TODO: giovanni to finish} describe CCTA
Our original control algorithm \cite{ICCD_bambini}, \cite{CCTA_bambini} is based on a cascade structure and it uses the cores' estimated power consumption as the controlled variable. Contrarily to ARM, the power budget is firstly distributed among cores with a heuristic algorithm based on the temperature state of each core, then this granted power is further reduced to meet the thermal set-point through a distributed PID layer.

Albeit being employed in fully-fledged modern \glspl{cpu}, such algorithms rely on control strategies that do not account for the general problem of control variable coupling. We believe that an analysis of this scenario is required in the research community, provided the recent availability of energy-efficient yet powerful parallel embedded microcontrollers that can sustain more complex control strategies.

%%%%%%%%%%%%%%%%%%%%%%
%TODO: Add a description of the \gls{llc} (On the other hand, the \glspl{llc} need to be tailored to the system they are controlling.)

%... Even with these limitations, ~\cite{survey1, survey2},  In these works, \gls{dvfs} typically relies on architectural simulators, or do not explicitly consider coupling between F and V, and their constraints.

%Often \gls{dpm} tecniques are influence by other factors than the power relation between F and V, such as the number of active cores, or scheduling the type of application

%Industry generally use frequency, while soA research use power  another difference between \gls{hlc} and \gls{llc}

\begin{table*}[t]
    \caption{Summary of existing \gls{hlc} and \gls{llc} \gls{sota} implementation in industry and academia. 
    The \dag~ symbol in the Step Time column indicates the presence of event-based controller features. VBA stands for Voting Box Algorithm.}
    \begin{center}
    \renewcommand{\arraystretch}{1.3} % Tune row height (vertical spacing)
    %\resizebox{\linewidth}{!}{
    %\begin{adjustbox}{width=\textwidth}
    \begin{tabular}{cc|cc|cc}

        \textbf{Reference} & \textbf{Year} & \textbf{Type} & \textbf{Target} & \textbf{\thead{Controller \\ type}} & \textbf{\thead{Step Time \\ in $ms$}} \\
        %\textbf{\thead{\gls{pcf} \\ scheduling}}
        
        \hline %\hline
        \multicolumn{6}{c}{\textbf{Academia}} \\ 
        \hline %\hline

        Bartolini et al.~\cite{TPDS} & 2013 & \gls{hlc} & HPC & MPC &  \\

        Bhat et al.~\cite{paperB} & 2017 & \gls{hlc} & Mobile & Model-Based & $100^{\dag}$ \\
        Cui et al.~\cite{fuzzy-variation} & 2017 & \gls{hlc} & Mobile & Fuzzy & $1-20$ \\
        Majumdar et al.~\cite{gpgpu} & 2017 & \gls{hlc} & Desktop / Mobile & MPC & \dag \\
        Rapp et al.~\cite{paperA} & 2017 & Boost & HPC & ML & $1$ \\

        Leva et al.~\cite{leva-event-based} & 2018 & \gls{hlc} & Desktop & PI & \dag \\
        
        Moazzemi et al.~\cite{hessle-free} & 2019 & \gls{hlc} & Mobile & Fuzzy & $200$ \\
        
        Mandal et al.~\cite{energy-aware} & 2020 & \gls{hlc} & Mobile & ML & $50$ \\

        Tilli et al.~\cite{Tilli2022} & 2022 & \gls{hlc} & HPC / Desktop & MPC &  \\
        Bambini et al.~\cite{ICCD_bambini} & 2022 & \gls{llc} & HPC & PI + Heuristic & $0.5$ \\
        \textbf{This work} & 2024 & \gls{llc} & HPC & Hybrid & $0.5$ \\

        \hline %\hline
        \multicolumn{6}{c}{\textbf{Industry}} \\ 
        \hline %\hline

        PCU (Intel) & in use & \gls{llc} & HPC / Desktop / Mobile & PI VBA & $0.5$ \\
        SCP, MCP (ARM)  & in use & \gls{llc} & HPC / Desktop / Mobile & PI VBA & n.d. \\
        SMU (AMD) & in use & \gls{llc} & HPC / Desktop / Mobile & PI VBA & n.d.  \\
        OCC (IBM) & in use & \gls{llc} & HPC & PI VBA & $0.25$ \\
        
        % Energy-Aware High-Performance Computing: Survey of State-of-the-Art Tools, Techniques, and Environments

        \hline
        \end{tabular}
    %}
    %\end{adjustbox}
    \label{tab:pcs_sota}
    \end{center}
\end{table*}
\section{Background}\label{sec:background}

%%% SubChapt: Mathematical Model

%%%%%%%%%%%%%%%%%%%%%%%%%%%%%
%%% Present the Structure of the HPC processor

%%il workload dal punto di vista del controllo è un disturbo, non è controllabile. 
As fundamental units of modern supercomputers, \gls{hpc} processors are nowadays heterogeneous systems comprising several computing chiplets connected with an \emph{interposer} by means of \glspl{tsv}~\cite{AMD_CHIPLET} and integrating on-chip HBM memories.% and \glspl{dsa} such as \glspl{pmca} and \glspl{ha}. %

In the realm of power and thermal management, a precise model capturing its physical dynamics is of paramount importance.
In this work, we focus on modeling a computing chiplet integrating the main application class \glspl{pe} as the most significant sources of power and temperature fluctuations.
From a control perspective, each \gls{pe} can be considered as a \gls{mimo} system where the supply voltage $V_{i}$ and the operation frequency $F_{i}$ are the controlled \emph{input} variables, while the temperature $T_{i}$ and the power consumption $P_{i}$ are the \emph{output} variables, as shown in ~\cref{fig:component}. \reb{Frequency and Voltage are linked by a functional inequality constraint, as described more in detail in \cref{ssec:realact}, but they retain a degree of independence.}
These four variables are the main elements of the control problem.

\begin{figure*}[t]
	\centering
	\includegraphics[width=0.6\linewidth]{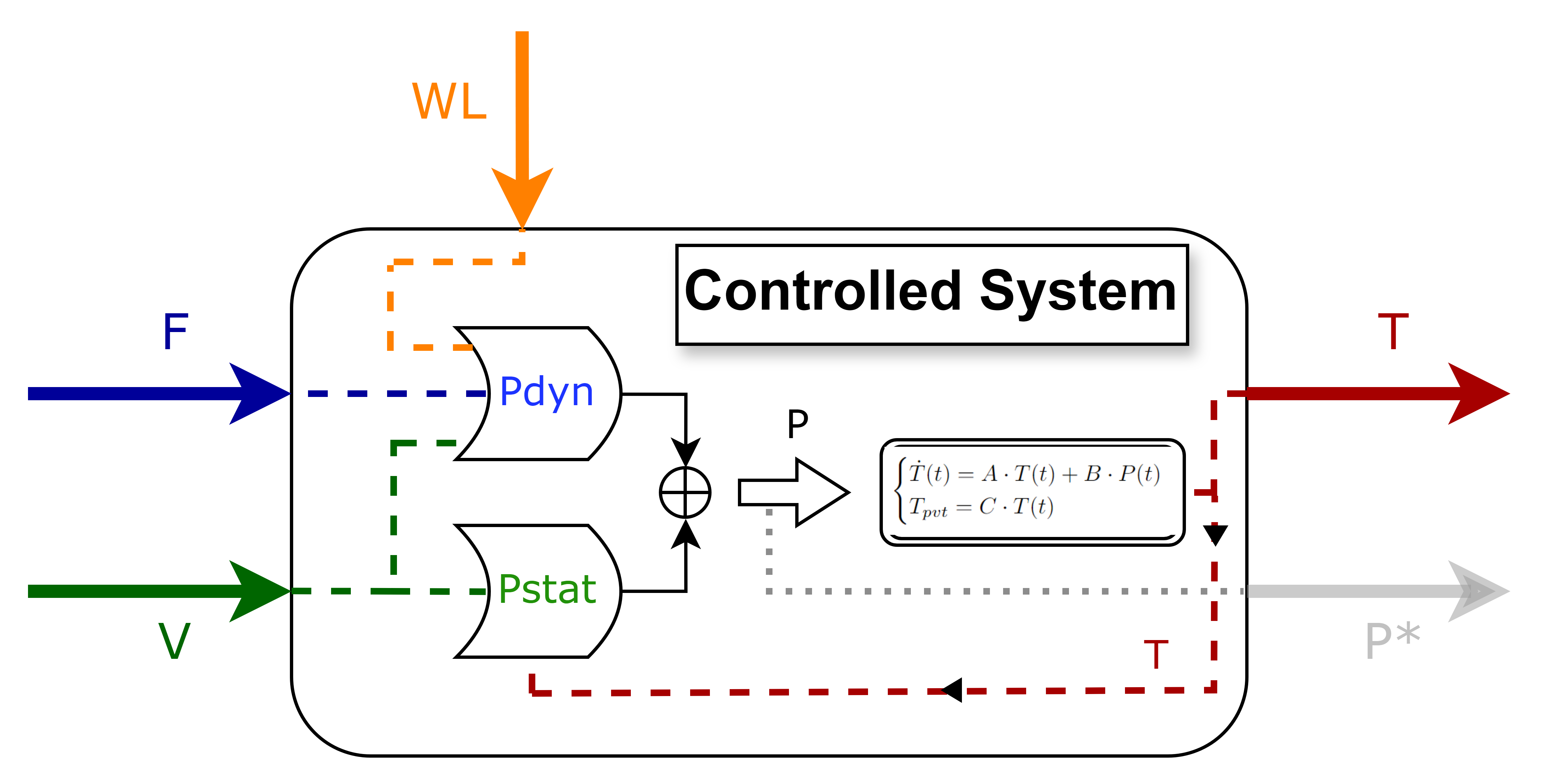}
	\caption{\label{fig:component} Representation of a \gls{pe}-component with the four main elements of the control problem (the inputs Frequency ($F$) and Voltage ($V$), and the outputs Temperature ($T$) and consumed Power ($P$))  and the workload $\omega_i(t)$. The power consumption $P*$ is grayed out because it is not directly measurable per single \gls{pe}. Instead, the power consumption measure is provided for each power rail (i.e. group of \glspl{pe}).}
	%\vspace{-15pt}
\end{figure*}

The power consumption $P_{i}(t)$ of each \gls{pe} component $c_i$ of the system can be described by an algebraic non-linear function that links the frequency $F_{i}(t)$, the voltage $V_{i}(t)$, and the temperature $T_{i}(t)$ with the workload $\omega_{i}(t)$ (i.e. the application) executing on the specific component itself. Different workloads activate different gates of the \gls{pe}, producing different power outputs. Frequency and workload scale linearly, voltage and temperature have a non-linear relationship~\cite{Hanumaiah, thermal_survey}. %
Differently from power consumption, the temperature $T_{i}$ of each component is a state of the system evolving in time subjected to $P_{i}(t)$ and nearby temperatures of neighboring \glspl{pe}~\cite{TPDS, Tilli2022}.

\Cref{fig:chip} shows the architecture we consider for deriving the mathematical description of the heat dissipation dynamic system.
It represents a chiplet integrated on a silicon die over a carrier \gls{pcb}, with a copper heat spreader placed over the active silicon devices to ease heat dissipation and facilitate the mounting of the aluminum heat sink.
\begin{figure*}
	\centering
	\includegraphics[width=0.6\linewidth]{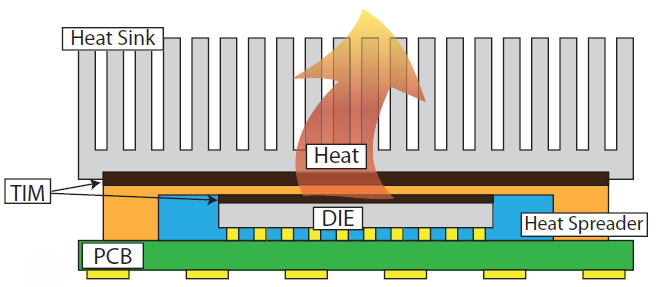}
	\caption{\label{fig:chip} Representation of the HPC Processor thermal architecture. The die contains all the \glspl{pe} which are the source of the heat production through power dissipation. The main heat dissipation path is indicated by the red arrow passing through the Heat Spreader, the Heat sink, and two TIMs layers. The secondary heat dissipation path goes through the \gls{pcb} layer below the die to the Motherboard.}
	%\vspace{-15pt}
\end{figure*}
From physics first principles and applying the Fourier heat-equation to the silicon and metallic layer, we obtain the following Partial Differential Equations (PDEs):
\begin{align}
\rho_{Si}c_{Si} \frac{\partial T_{Si}(x, t)}{\partial t} &= k_{Si}\nabla^2T_{Si}(x, t) + q(x, t),  \\ 
&\qquad \text{with:} \; \; x \in V_{Si}, t \in \mathbb{R}_{\geq0}\label{eq:Si_heat} \\
%& \notag \\
\rho_{Cu}c_{Cu} \frac{\partial T_{Cu}(x, t)}{\partial t} &= k_{Cu}\nabla^2T_{Cu}(x, t), \\
& \qquad
 \text{with:} \; \; x \in V_{Cu}, t \in \mathbb{R}_{\geq0},\label{eq:Cu_heat}
\end{align}
%with (Dirichelet) boundary, and (time) initial conditions
%\begin{equation}\label{eq:bl_and_ic}
%\begin{aligned}
%&T_{Si}(x, 0) = T_{Si, 0}(x), \qquad \forall x \in V_{Si}\\
%&T_{Si}(x, t) = T_{\partial Si}(x, t), \qquad \forall x \in \partial V_{Si}\backslash\partial V_{Cu}\\
%&T_{Cu}(x, 0) = T_{Cu, 0}(x), \qquad \forall x \in V_{Cu}\\
%&T_{Cu}(x, t) = T_{\partial Cu}(x, t), \qquad \forall x \in \partial V_{Cu}\backslash\partial V_{Si}\\
%&T_{Si}(x, t) = T_{Cu}(x, t) \\
%&k_{Si} \nabla T_{Si}(x, t) = k_{Cu} \nabla T_{Cu}(x, t), \\
%& \qquad \forall x \in \partial V_{Si} \cap \partial V_{Cu}, t \in \mathbb{R}_{\geq0},
%\end{aligned}
%\end{equation}
%
where $T_{Si}(\cdot)$ and $T_{Cu}(\cdot)$ are the temperatures of the silicon device and the heat spreader defined in the open volumes $V_{Si}$, $V_{Cu}$, $q(\cdot) \geq 0$ is volumetric thermal power generated by internal sources (e.g. cores power outputs), and $\rho_{Si}$, $\rho_{Cu}$, $c_{Si}$, $c_{Cu}$, $k_{Si}$, $k_{Cu}$ are the density, specific heat and thermal conductivity of the two materials respectively. 

An approximated model of the aluminum heat sink and the \gls{pcb} is presented in \cref{ssec:addthermmodel} as an extension to the model of the cores and the heat spreader since (i) the time-constants for the heat sink and the \gls{pcb} are three to four order of magnitude slower than the faster thermal time constant of the Silicon~\cite{hotspotv1} making it less relevant from the controller point of view, (ii) the heat sink thermal models is highly variable, depending on its shape, characteristics, and the variable airflow of the fans, and (iii) the \gls{pcb} is can not be modeled and the lower surface of the cores can be considered adiabatic~\cite{TPDS, thermal_survey}. Thus, modeling them with the same precision as the cores will introduce unnecessary complexity in the thermal model. For the same reasons, the model of the interposer layer, required by the chiplet design, is merged with the \gls{pcb} model.

%
%\section{Control Model Definition}\label{ssec:control_mod_def}
%
\begin{figure*}%[h]
	\centering
	\includegraphics[width=0.7\linewidth]{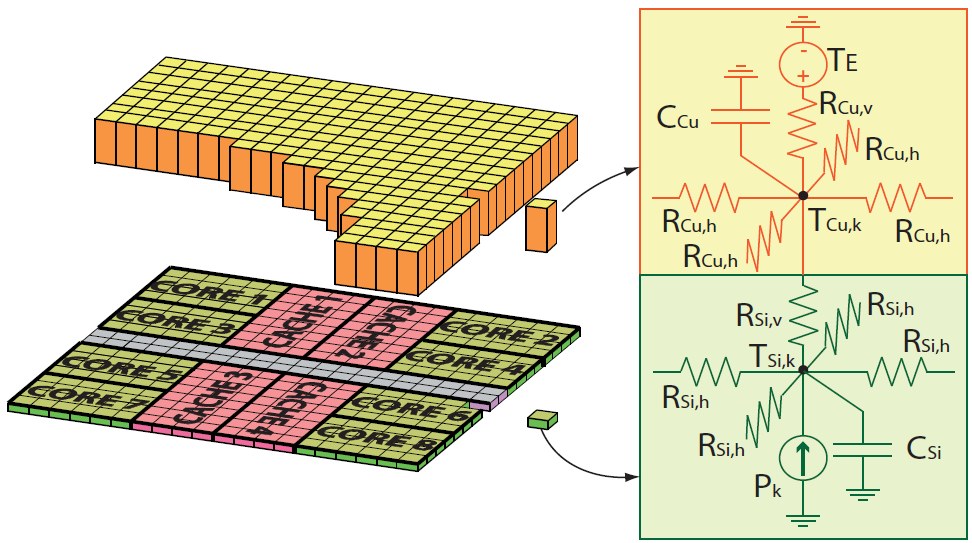}
	\caption{\label{fig:thermal_fem}Example of finite elements spatial discretization, with corresponding lumped parameters model of a single \gls{pe}. In the model, the green part (below) concerns the core, with $P_k$ being the consumed power generated by the \gls{pe}, the yellow part (above) concerns the heat-spreader, with $T_E$ being the main dissipation path to the Heat Sink.}
	%\vspace{-20pt}
\end{figure*}

To obtain a model which can be tractable for control design purposes (control model) the PDEs in \eqref{eq:Si_heat}-\eqref{eq:Cu_heat} can be ``converted'' into Ordinary Differential Equations (ODEs) by employing Euler discretization of the derivatives with respect to the spatial coordinates $x$. Clearly, this operation introduces some degree of approximation. 
However, resorting to the structural properties of thermal systems, for which the \emph{Maximum principle}~\cite{Evans}
holds, it can be shown~\cite{Tilli2022, TPDS, hotspotv1} that the feasibility of the thermal constraints can be guaranteed by discretizing the original PDEs in space and focusing on the hot spots of the die.
Therefore, for simplicity we select $n_{c}$ finite elements (cells) each associated with a power source, i.e. a \gls{pe}. 
Multiple elements could be assigned to each power source with a slight increase in the notation complexity, similar to what is shown in \cref{fig:thermal_fem}.
Note that each finite element has two thermal states, one associated with the (local) temperature in the silicon die and the other with the local temperature of the metallic heat spreader.
The power source $P_i$ associated to each source is described as follows:
\begin{equation}\label{eq:pow_map}
\begin{aligned}
P_i(t) = \int_{V_{s, i}}q(x,t)dV = h_i(F_i(t),V_{i}(t), \omega_i(t),T_{Si,i}(t)),
\end{aligned}
\end{equation}
where $h_i$ is the nonlinear mapping function. % that we assume to be invertible in an open connected interval $\mathcal{F}$. %(i.e., given the power consumption and knowing the source workload and temperature, we can always find a unique pair of frequency and voltages producing that power).
Different models can be used for $h_i$ without significant modifications to the proposed \gls{hpc} processor model. The power model \eqref{eq:act_pow_map} employed in this work is 
%an approximated version of~\cite{TPDS} which is 
the most commonly used one~\cite{thermal_survey}. %, where the $k$ parameters are constant and computed on the critical values of $V_{\text{MAX}}$ and $T_{\text{MAX}}$, and $\omega_L$ capture the effective capacitance relation to the executed workload.
\begin{equation}\label{eq:act_pow_map}
\begin{aligned}
P_{i} &= P_{i,\text{stat}} + P_{i,\text{dyn}} \\
&= k_{s_0} + (I_{cc,i} V_i) \cdot \mathcal{K}(T_{Si,i}, V_i) + C_{eff,i} F_i V_{i}^2
\end{aligned}
\end{equation}
%%%%%%%%%%%%%%%%%%%%%%%
%
where $P_{i,\text{stat}}$, $P_{i,\text{dyn}}$ are the static and dynamic component of the \gls{pe}s' power consumption, depending on the static current $I_{\text{cc,i}}$, and the effective capacitance $C_{\text{eff,i}}$ respectively.
$C_{eff}$ is correlated to the type of instructions ($\omega_i(t)$) being executed on the \gls{pe}. $\mathcal{K}(T_{Si,i}, V_i)$ is a non-linear mapping that encapsulates the dependency of the static leakage power to the temperature and the voltage of the component. 

Adding the temperature dependency to the power model (i) raises the complexity of the control algorithm since both the temperature $T_i$ and the power consumption $P_i$ of each component are tightly coupled between each other, (ii) creates scenarios where constant inputs ($F_i,V_{i}$) and workload $\omega_i$ produce variation in the power consumption $P_{i}$, and (iii) models the thermal run-away scenario \cite{thermal_runaway} where temperature and power increase each other in a positive feedback, leading to the irreversible damage of the component.

In this work an exponential relation based on \cite{bartolini2019advances, ROSSI2016} is employed:

%In this work an exponential relation based on \cite{mosfet_exp} is employed:
\begin{equation}\label{eq:exp_leakage}
\mathcal{K}(T_{Si,i}, V_i) = e^{\bigl(k_v V_i(t) + k_T T_i(t) + k_{T_0} \bigr)}
\end{equation}
%\textcolor{red}{TODO @Andrea: }
where the $k$ parameters are constant and computed on the critical values of $V_{\text{MAX}}$ and $T_{\text{MAX}}$.

Denoting with $\mathcal{N}\{i\}$ the set of all neighbors of the element $i$, then the differential equation associated with a generic element reads as:
\begin{equation} 
\label{eq:lump-par-mod}
\begin{split}
\dot T_{Si,i} &= \frac{P_{i}}{C_{Si, i}} + \frac{T_{Cu,i}-T_{Si,i}}{C_{Si, i} R_{Si, i}^v}  +  \sum_{j \in \mathcal{N}\{ i\}}\frac{T_{Si, j}-T_{Si,i}}{C_{Si, i} R_{Si,ij}^h}\\
\dot T_{Cu,i} &= \frac{T_{Si, i}-T_{Cu, i}}{C_{Cu, i}
	R_{Si, i}^v}  +  \frac{T_{{Al}} - T_{Cu,i}}{C_{Si, i} R_{Cu,i}^v}  + \sum_{j \in \mathcal{N}\{ i\}}\frac{T_{Cu,j}-T_{Cu,i}}{
	C_{Cu, i} R_{Cu, ij}^h},
\end{split}
\end{equation}
where $C_{Si, i}$, $C_{Cu, i}$ are the thermal capacitances, while $R_{Si, i}^v$, $R_{Cu,i}^v$, $R_{Si, ij}^h$, $R_{Cu, ij}^h$, $j \in \mathcal{N}\{ i\}$ are respectively the vertical and horizontal thermal resistances, and $T_{{Al}}$ is the temperature of the aluminum heat sink. These lumped parameters stem from the aforementioned spatial discretization procedure as shown more in detail in~\cite{paci_thermal_rc} and~\cite{hotspotv1}.
\begin{equation} 
\label{eq:lump-par-RC}
\begin{split}
C &= c^{Si/Cu} \cdot h \cdot w \cdot l \\
R^{v} &= k^{Si/Cu} \cdot \frac{h \cdot w}{l} \\
R^h &= k^{Si/Cu} \cdot \frac{l \cdot w}{h}
\end{split}
\end{equation}
%

%For simplicity, in the following we consider $n_c = n_s$.
%Indeed, if $n_c > n_s$ the power of each source can be assumed to be uniformly split among the cells that are fed by that source.
%For instance, if $n_c = M n_s$, and for each source $j$ are associated $M$ finite elements, then the $i$-th power of each of those $M$ elements is given by $P_i = P_{s, j}/M$, where $P_{s, j}$ is computed as in \eqref{eq:pow_map}.
Collecting all temperatures in a unique vector $T=(T_{Si,1}, T_{Cu,1}, \ldots, T_{Si,n_s}, T_{Cu,n_s}, T_{{Al}})^T$  we can compactly rewrite system \eqref{eq:lump-par-mod} as:
%
%\begin{equation}\label{eq:lumped_par_mod_comp}
%\dot{T}=A T + BP, \qquad T_{Si}=CT,
%\end{equation}
\begin{equation}\label{eq:lumped_par_mod_comp}
\begin{cases}
\dot{T}(t)= \mathbf{A} T(t) + \mathbf{B} P(t)  \\
 T_{Si}= \mathbf{C} T(t)
\end{cases}
\end{equation}
where $P = (P_1, \ldots, P_{n_c})^T$, and $\mathbf{A}$, $\mathbf{B}$, $\mathbf{C}$ follow directly. Dynamics in \eqref{eq:lumped_par_mod_comp} combined with the algebraic power model in \eqref{eq:act_pow_map} describe the overall system thermal and power behavior.

\subsection{Thermal Model extensions} \label{ssec:addthermmodel}

The modeling of the aluminum heat sink and \gls{pcb} is realized by adding a state in the system representing each entire layer. These states exchange heat with the connected blocks through heat conduction, with the addition of a spreading resistance in series to account for the different contact areas, as shown in~\cite{hotspotv1} following the formula in \cite{spreadingR}. %
Albeit being an approximation~\cite{riva2021numerical}, with this approach we aim at introducing slow time-varying temperature and dissipation shifts to study the impact on the controllers.

To complete the thermal dissipation path, two additional states for the motherboard and the air are added to the thermal model, as well as additional resistances to emulate the presence of the two different \glspl{tim} between the cores and the heat spreader, and the latter and the heat sink. The vectors and matrices of the model \eqref{eq:lumped_par_mod_comp} are extended with the inclusion of the described four states.

%\textcolor{red}{TODO:} According to \cite{[hottypotty]} and the spatial discretization approximation based on on the hot spots of the die presented above, it is possible to introduce thermal coupling with on-die lateral memory controllers and with adjacent chiplets, treating them as large power producing cells employing the \textit{constriction resistance}.

%\textcolor{red}{TODO:} as of \cite{[non riesco a ritrovare il paper]}, thermal variations influence also the value of the $R^{Si/Cu}$ and $C^{Si/Cu}$ parameters introducing further non-linearities

\subsection{Modeling Real Actuators}\label{ssec:realact}

As shown in \cref{fig:component}, the control outputs are the Frequency $F$ and the Voltage $V$. Most of the control challenges proposed in this work derive from the non-idealities of these two actuators. %Hence, it is crucial to model them correctly to avoid results marred by oversimplification. 

The Frequency and Voltage variables are not independent. A \gls{fet}-based digital circuit requires continuously toggling the \glspl{fet}'s state, changing its gate voltage below or above a certain threshold. The rate of speed at which this change can happen (the device \emph{switching speed}) contributes to the maximum frequency at which that digital circuit can run operations. Toggling the \gls{fet} by changing its gate voltage requires charging or discharging its transistor capacitance. %
Higher voltages result in a faster slew rate when charging and discharging, allowing an increased switching speed. Additionally, the further away the gate voltage from the threshold voltage, the lower the resistance of the \gls{fet}'s conducting channel, resulting in a lower RC time constant and thus a quicker realization of the logic state. %
For these physical reasons, maximum frequencies are bounded by the currently supplying voltage. Reaching higher frequencies requires increasing the voltage, and the relation between Frequency and Voltage is sub-linear~\cite{digitalbook1,bartolini2019advances}.

Another non-ideality that is common in multi-cores processors design is that Frequency and Voltage actuators may be shared among several \gls{pe}s. This is often the case for the Voltage, which is shared among several cores that are thus divided into \emph{Voltage Domains}, while it is more common to find \gls{pe}s with exclusive PLLs or frequency dividers. Requiring multiple cores to share the same voltage introduces a coupling effect in the output of the controller, which may result in degraded performance.

Indicating with $\mathcal{D}_j$ the $j^{\text{th}}$ power rail that supplies the component $c_i$, and omitting the time dependency for ease of reading, we can describe the two above restrictions as:

\begin{equation} \label{eq:FV}
    F_{i} \in \; [ \, F_{\text{min}}^S, \; F_{\text{max}}(V_{\mathcal{D}_j}, T_i) \,] 
         \qquad \text{with } \; c_i \in \mathcal{D}_j
\end{equation} 
where $F_{\text{min}}^S$ is the minimum core frequency of the system.

Frequency and Voltage are typically generated by \glspl{pll}, \glspl{fll}, or frequency dividers and \glspl{vrm}, respectively. These components generate discrete values and introduce delays (\emph{transition times}) when changing from one discrete value to the other, requiring either to halt the processor or to keep a minimum value for the transition time, decreasing the application execution performance~\cite{FV_delay}.

In this paper we model these delays with the following assumptions: (i) we assume two \glspl{pll} for each frequency output, thus when one is changing to a new value the other can maintain the current frequency without halting the processor, (ii) voltage transitions require waiting at a saturated frequency the completion of the transition, thus we can assume a constant maximum delayed response in the changed output. \reb{Although the voltage transition is completed in less than $50 \mu s$ which is faster than the control loop~\cite{FV_delay}, it is essential to model these delays as they add to the whole controller response latency, allowing us to study more in detail the impact of the control step periodicity on the control performance (see \cref{sssec:ctrl_challenges:delay} - Controller delays).}

%%%%%%%%%%%%%%%%%%%%%%%%%%%%%%%%%%%%%%%%%%%%%%
%%%%%%%%%%%%%%%%%%%%%%%%%%%%%%%%%%%%%%%%%%%%%%
\subsection{Exponential Leakage Analysis}\label{ssec:expleak}

\begin{figure*}[h!]
	\centering
	\includegraphics[width=1\linewidth]{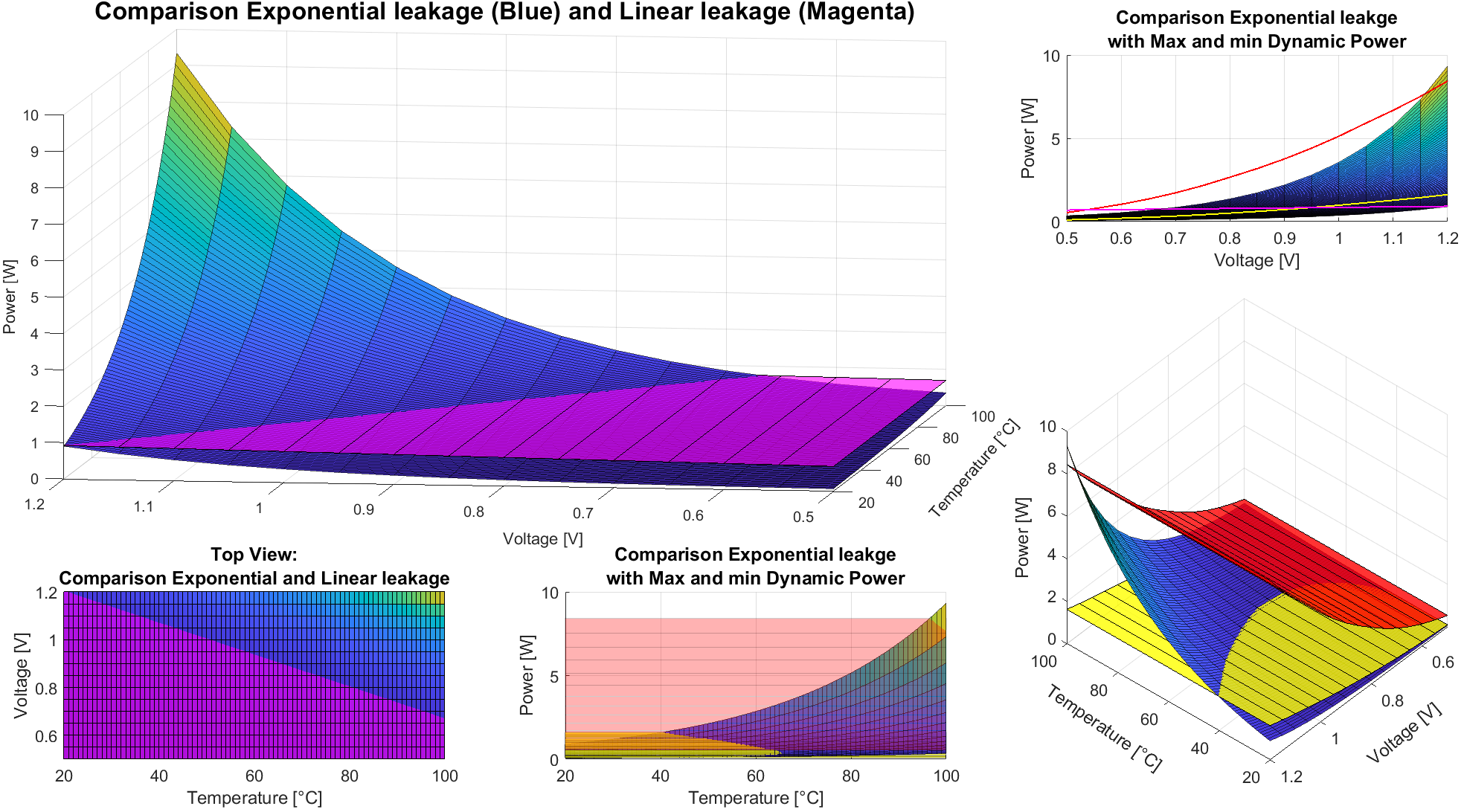}
	\caption{\label{fig:leakage} Multi-view comparison between the proposed model of the leakage power characterized by an exponential relationship with Voltage and Temperature (blue surface), the same model of leakage power with no exponential relation (magenta surface), and the minimum and maximum dynamic power (yellow and red surface respectively).}
	%\vspace{-15pt}
\end{figure*}

%As described in \cref{sec:background}, in this paper we introduce an exponential leakage model. 
\Cref{fig:leakage} shows the comparison between the exponential leakage power model (EXP-LP) \eqref{eq:exp_leakage} and the same model without exponential relation (L-LP). In the first plot, we observe the interval of values that characterize the exponential relationship of EXP-LP to $V$ and $T$, and the gap with the flatter magenta plane (L-LP model), with a power consumption that can be up to 10 times higher at high voltages and temperatures. The top view of the graph in the bottom left shows that for most operating points the EXP-LP model power consumption is less than or equal to the L-LP model, thus the exponential behavior of the static leakage power shows at operating points that are compatible with high-performance scenarios.

In the other plots, we can observe that at those operating points, the leakage power becomes a relevant or even predominant fraction of the total power consumption. 

%\reb{TODO: expand this part with consideration of the impact of this exponential relation model to the control performance }

%%%%%%%%%%%%%%%%%%%%%%%%%%%%%%%%%%%%%%%%%%%%%%
%%%%%%%%%%%%%%%%%%%%%%%%%%%%%%%%%%%%%%%%%%%%%%
\subsection{Model Implementation in Simulation}\label{sec:model_impl}
%If we consider the power model \eqref{eq:act_pow_map} as a part of the HPC processor model with \eqref{eq:lumped_par_mod_comp}, the Core Instantaneous Power Consumption input to the model is replaced by another type of input, depending on the model itself. In this paper, the input is the type of executed instruction.
%In the implementation the HPC processor model has been divided into three virtual cascade blocks: the first one (\textbf{Input Block}) takes the simulation input sequences and constants, arranging them to be fed as inputs for the subsequent blocks, the second one (\textbf{Power Model Block}) computes the Instantaneous Power Consumption of each core through the power mapping \eqref{eq:act_pow_map}, and the third one (\textbf{Thermal Model Block}) computes the thermal model based on equations \eqref{eq:lumped_par_mod_comp}.

%For performance reasons, instead of adopting a continuous simulation, we decided to discretize the thermal model \eqref{eq:lumped_par_mod_comp} of the third block with a step (Model Step) that is $1/1000$ the step of the control loop. 

%Instead of adopting a continuous simulation, we decided to discretize the state-space model \eqref{eq:lumped_par_mod_comp} with a step (Model Step) of $500 ns$.
%, which is $1/1000$ of the interval of the control loop. 
There are two different timing requirements to run the presented model in a simulation: the timing restriction related to the power model \eqref{eq:act_pow_map} and the one related to the thermal model \eqref{eq:lumped_par_mod_comp}.
According to~\cite{Beneventi} the thermal timing constants inside a processor are $\sim10s$, $\sim0.1s$, $\sim1ms$. The power consumption of each component $P_i(t)$ varies according to the workload $\omega_i(t)$ and the frequency of operation $F_i(t)$. Assuming the frequency to be in the GHz scale, and that the workload could be constantly changing, the power consumption $P_i(t)$ would be changing with timings below the $ns$ scale. 

Running a simulation with $ns$ steps would be superfluous for two reasons: power spikes between $10us$ and $100ns$ are mitigated by hardware elements of the \gls{pdn}~\cite{PDN}, and it is extremely difficult to obtain instructions sequences with a granularity of $ns$ since processors inspection tools provided by the manufacturer (as event and performance counters) have generally a sampling time of $1ms$\footnote{1ms is the default tick period of the operating system scheduler, which makes performance monitoring less intrusive if done synchronously with it \cite{bartolini2019advances, schone_energy_2019}.}~\cite{rapl_2015}. %
Choosing a simulation step $t_s$ that is slower than the frequency $F$, requires to use an average power consumption $\overline{P_{i}}(t_s)$ instead of the instantaneous one.

Introducing $\mathcal{S}_i = ( C_0, C_1, ... )$ as the sequence of \textit{instructions-characterized cycles} to be executed on the i-th core\footnote{Consider this as a simulation abstraction since the instructions sequence to be executed and the IPC are not generally known a priori.}, each element of the sequence corresponds to 1 cycle, thus describing not only the instructions sequence but also characterizing instructions IPC, memory stalls, cache misses, SIMD/Vector instructions, etc.
During each simulation step $t_s$, assuming the core frequency $F_i(t)$ and the voltage $V_i(t)$ to be constant, a variable number of instructions $\theta_{v_i}(t_s)$ from $\mathcal{S}_i$ will be executed, with $\theta_{v_i}(t_s) = F_i(t_s) \cdot t_s$.
Assuming the simulation step $t_s$ to be appropriate orders of magnitude faster than the fastest thermal time constant, in each $t_s$ only the $C_{eff,i}$ component of the model \eqref{eq:act_pow_map} will vary depending on the type of workload. Thus computing an effective capacitance density across each $t_s$, would allow us to compute $\overline{P_{i}}(t_s)$.

\begin{equation}
\begin{aligned}
    \overline{P_{i}}(t_s) &=  k_{s_0} + (I_{cc,i} V_i) \cdot \mathcal{K}(T_{Si,i}, V_i) +  \overline{C_{eff,i}} F_i V_{i}^2 \\
    & \quad \text{with:} \quad \overline{C_{eff,i}} = \frac{\sum_{i=1}^{\theta_{v_i}(t_s)} C_{eff,i}(t)}{\theta_{v_i}(t_s)} \\
\end{aligned}
\end{equation}

%\textcolor{red}{TODO: to avoid overnotation, $P_{i}(t)$ will be used as $\overline{P_{i}}(t_s)$ }

We chose a simulation step of $50 us$, which is two orders of magnitude faster than the fastest timing constants and in the range where the Power Delivery Network mitigates power spikes. %
We introduce parameter variation in the coefficients of the power model \eqref{eq:act_pow_map} and thermal model \eqref{eq:lumped_par_mod_comp} to encapsulate the manufacturing difference and process variation from core to core~\cite{bartolini2019advances, FINFET_PVT}. 
%with the noise we try to add small variations in the power consumption since the instruction sequence we use is `artificial' and block-based, instead of in reality every instruction is generally different from the previous one.

Finally, we model white noise at the output of the thermal model $T_{Si}$ to emulate the cores \gls{pvt} sensors~\cite{TCAS}.

%The Input Block will fetch the sequence of instructions and provide, for each Model Step and each core, an `\textit{instructions density input}' to the Power Model Block to compute the Average Power Consumption. Note that the fetched instructions sequence dimension depends on the clock frequency at which the core is running in the Step. 

\section{Control Problem}\label{sec:ctrlobj}

%%%%%%%%%%%%%%%%%%%%%%%%%%%%%
%%% Present and derive control requirements and timings
\subsection{Formulation}\label{ssec:ctrl_formulation}

From a thermal perspective, each component of a \gls{cpu} has a critical thermal limit $T_{\text{CRIT}}$ causing irreversible damage if exceeded. 
High temperatures lead to faster aging as well as performance and reliability degradation~\cite{AGING_1}, modeled through a thermal limit $T_{\text{L}}$. Compared to $T_{\text{CRIT}}$, $T_{\text{L}}$ is a softer restriction that, if exceeded temporarily and infrequently, does not cause damage.

Another type of physical limitation is given by the maximum power that can be supplied to the processor. 
As the effects of Dennard scaling came to an end in the first decade of 2000, the power density constantly increased at each technology scaling, limiting the power and the current in today's processors. 
Pads in the package are shared among power delivery networks, memories, and I/O interfaces, generating a trade-off between peak current and data bandwidth. %
At the board level, \glspl{vrm} capacitors and inductors become dominant. As a result, modern processors' clock frequency is bounded by the power that can be provided to the processor itself in a given instant and not always by its absolute and attainable physical speed.
%Since the slowdown of Moore's law in the first decade of 2000, an increasing number of \glspl{pe} with an overall peak power consumption greater than the maximum sustainable power has been packed in a single processor to exploit parallel and different-load phases execution~\cite{[]}. 
%
For this reason, system-level and voltage-domain-level power consumption limits need to be enforced. %Generally \cite{[]}, this power limitation does not involve single elements of the processor since the parts are designed to sustain their maximum reachable load singularly.

As described in \cref{sec:background}, the temperature $T_i$ and the power consumption $P_i$ of a component are coupled in a positive feedback, as they increase when each other increases. 
High $P_i$ and/or $T_i$ cannot be sustained for a long period of time, as they can lead to an uncontrollable system condition referred to as \emph{thermal runaway}~\cite{thermal_runaway}. This detrimental behavior is caused by the exponential relationship between the leakage power and the temperature, as described in \cref{ssec:expleak}.

In addition to the thermal and power limitations due to the physical nature of the components, there are additional time-variable power and thermal limits in the form of requests from external agents, such as off-chip board and node management controllers, or the \gls{os}.
Assuming the requested limits are lower than the physical ones, we consider only the former in the control problem formulation. 
%the minimum request for each limit: for $T_{\text{L}}$, $P_{\text{B}_c}$ (power budget chip), and $P_{\text{D}_j}$ (power budget for each j-domain ).
%TODO finish the sentence

The performance objectives of the control algorithm are (i) the time of completion $t_{\text{exec}}$ of the application running on the \gls{hpc} chip, and (ii) the energy efficiency as in the energy $E_{\text{tot}}$ required to complete the execution~\cite{bartolini2019advances, HPC-survey19}. $E_{\text{tot}}$ can be computed as the integral of $P_T(t)$ over $t_{\text{exec}}$. 
$P_T(t)$ is the sum of all elements of the chip consuming power. Note that from model \eqref{eq:act_pow_map}, $P_T(t)$ has a static and a dynamic contribution. Hence, the execution speed plays a role in the energy efficiency. $t_{\text{exec}}$ depends on the \gls{pe}s running frequency, and its \gls{ipc}.

%%%%%%%%%%%%%%%%%%%%%%%%%%%%%
%%% Present HLC and LLC control problem formulation
Both the performance objectives described above depend on an exhaustive knowledge of the running application. For security and privacy concerns, the integrated \gls{llc} has no privileged access over the latter. Additionally, energy savings depend on other external components (\gls{gpu}, cooling system, power supply, ...) over which the \gls{llc} has no control, and, if the application execution is spread on different systems, the \gls{llc} has no global access on those systems too. 
The choice of the optimal operating point and the desired trade-off between $t_{\text{exec}}$ and $E_{\text{tot}}$ is therefore left to the \glspl{hlc}. 
The \gls{llc}'s objective is to apply the received operating point while enforcing the physical and requested thermal and power constraint described above within the required timing restrictions. 
Being implemented as a dedicated unit tightly coupled to the system under control, the \gls{llc} can perform low-level sensing and actuation at a faster pace than \glspl{hlc}.
Reductions in the operating points performed by the \gls{llc} due to thermal and power capping should still reflect the performance objectives dictated by the \glspl{hlc}. 

Provided the description above, the control problem is formulated as follows:
%
	%COMMENTED AND ADDED THERMAL CONSTRAINT
%\begin{equation} \label{eq:purp}
%\begin{aligned}
%& \min_{F_{a}} \sum_{k=0}^{N-1} \Big| F_{T}(t_k) - F_{a}(t_k) \Big|^2_R  \qquad i = 1, \ldots, n_c \\
%%
%& \text{min}\sum_{k=0}^{N-1} P_{tot}(t_k)
%\end{aligned}
%\end{equation}
%
\begin{equation} \label{eq:purp}
	\begin{aligned}
	 &\min_{F_{a}} \Big| F_{T}(t_k) - F_{a}(t_k) \Big|^2_{\mathcal{R}}  \\
	 &\text{subject to}: \ \sum_{i=1}^{n_c}P_i(t_k) \leq P_{\text{B}}(t_k) \\
	 &\qquad \qquad \quad \sum_{i=1}^{n_j}P_i(t_k) \leq P_{\mathcal{D}_j}(t_k), \quad i \in \mathcal{D}_j, \quad j=1,\dots ,  n_d \\
      &\qquad \qquad \quad T_i(t_k) \leq T_{\text{L}}(t_k) < T_{\text{CRIT}}, \quad i=1,\dots ,  n_c
	\end{aligned}
\end{equation}
%
%$N$ is the number of steps assumed as the horizon, $f=(f_{1},\,\dots,\,f_{n_s})^T$ represent the decision variables of the problem, i.e. the frequencies to be assigned to the cores by solving the problem above with the set of target frequencies $f_{t}=(f_{t,1},\ldots,f_{t,n_s})^T$
%and $R\in \mathbb{R}^{n_s \times n_s}$ is a symmetric positive definite weight matrix and $|\cdot|_R$ is the corresponding norm.
%
where $F_{a}=(F_{a1},\,\dots,\,F_{an_c})^T$ are the applied Frequencies to the $n_c$ cores, $F_{T} =(F_{T1},\ldots,F_{Tn_c})^T$ are the given Target Frequencies as operating point,
%$t_k$ are the discrete time instants, $N$ is the number of steps assumed as the horizon, 
and $\mathcal{R} \in \mathbb{R}^{n_c \times n_c}$ is a symmetric positive definite weight matrix and $|\cdot|_R$ is the corresponding norm. $P_{\text{B}}$ is the requested total power budget, $n_d$ is the number of the $\mathcal{D}_j$ voltage domains such that $\sum_{j=1}^{n_d}n_j = n_c$. Finally, $T_i$ is the $i^{th}$ core temperature.

The computation of the total power consumption as the sum of the power consumption over all the \gls{pe}s (\ref{eq:purp}.2) is not fully accurate as it neglects other entities such as \glspl{dram}, on-chip \glspl{hbm}, I/Os controllers, and the case of multiple chiplets. In this work, we focus on the \glspl{pe} control in a single chiplet scenario only. %Future works will extend the notation to consider 

%%%%%%%%%%%%%%%%%%%%%%%%%%%%%
%% Present control challenges
\subsection{Challenges}\label{ssec:ctrl_challenges}

\begin{figure*}[t]
	\centering
	\includegraphics[width=1\linewidth]{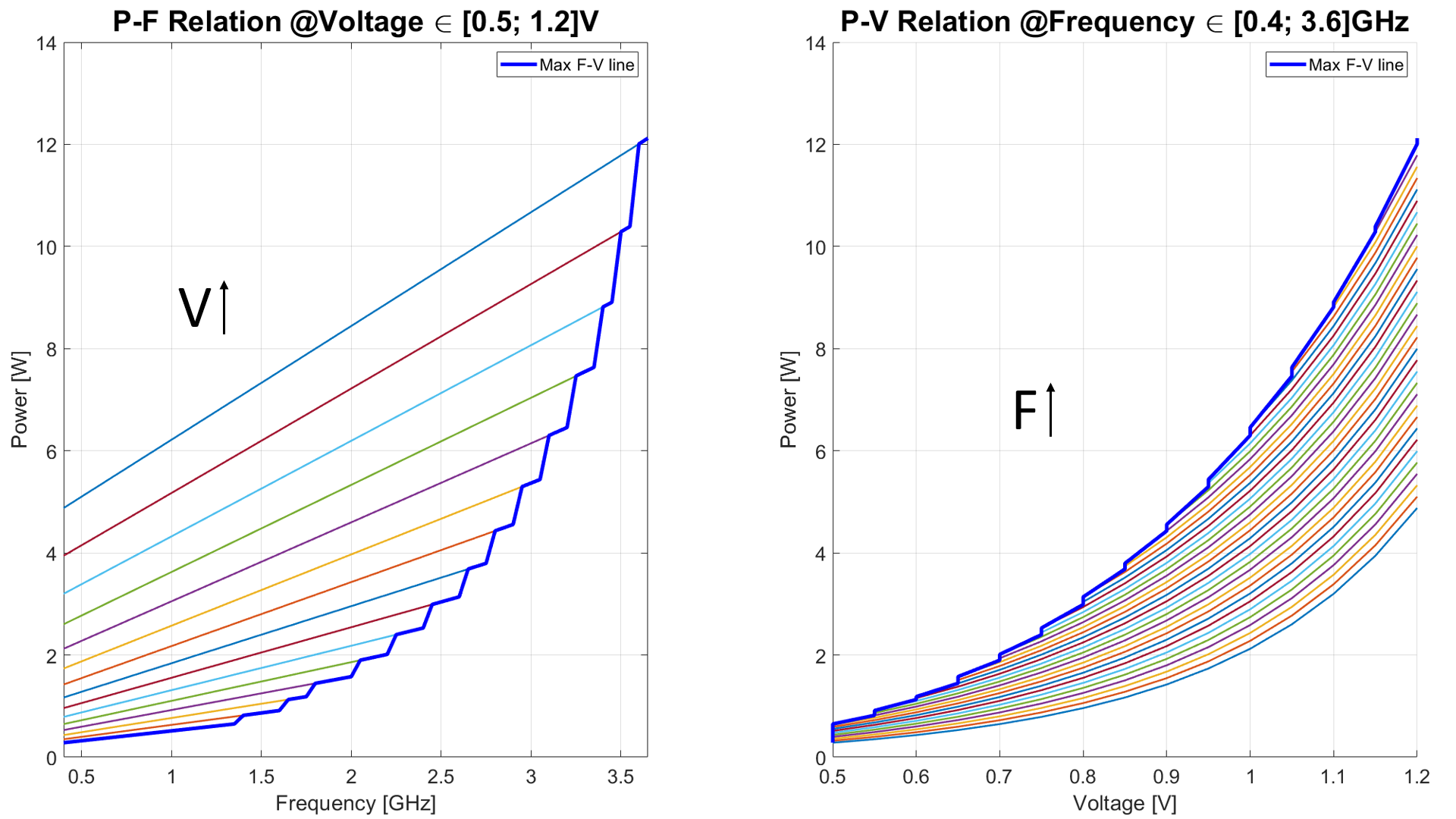}
	\caption{\label{fig:fvrel} The \gls{pe} power consumption $P(t)$ according to the model \eqref{eq:act_pow_map} in relation to frequency $F$ (left) and the voltage $V$ (right) at $75 \degree \text{C}$. Each graph only shows operating points that are feasible w.r.t. \eqref{eq:FV}. Different lines represent the non-plotted variable, i.e. the voltage in the left graph, and the frequency in the right one.}
	%\vspace{-15pt}
\end{figure*}

\reb{There are numerous state-of-the-art solutions to address the control problem \eqref{eq:purp}~\cite{PWR_THERMAL_CONSTRAINTS, Zhang2, HPC-survey19}; however, many of these solutions do not fully account for all the control challenges involved.}

First, the \gls{mimo} nature of the problem and the non-linearity of the controlled system, as shown in equations \eqref{eq:act_pow_map} and \eqref{eq:exp_leakage}. Then, the tightly coupled control objectives, the coupling, constraints, and discretization of the control outputs, and the high-amplitude and high-frequency unpredictable noise originated from the unknown workload $\omega(t)$. 
Finally, the algorithm has to be simple enough to execute with a frequency in the order of \reb{$2-10 kHz$ on an embedded microcontroller.
This frequency accounts for both the requirements given by the limitation of the \gls{pdn} and the faster thermal time constant (as described in \cref{sec:model_impl} and in \cref{sssec:ctrl_challenges:delay} - Controller delays), allowing a consistent control of the temperature and power capping according to the most used control rules \cite{discrete_book}.}

\subsubsection{Controller outputs}\label{sssec:ctrl_challenges:var_coupling}
\Cref{fig:fvrel} shows the progression of the power output of a \gls{pe}, according to the model of a real actuator. In the left figure, we observe that, except for the initial values, the power gap between different voltage levels is larger than the line slope along frequency values. The power gap along the maximum $F-V$ line (blue) increases by more than $3\times$ at $2.2 GHz$ and $5\times$ at $2.7 GHz$. 
This means that, once a voltage level is chosen, changing frequency within that level (up to the maximum allowed frequency relative to that voltage level \cref{ssec:realact}) generates little power (and thus thermal) differences. Instead, changing the voltage level causes larger power and thermal deviations. For example, along the max $F-V$ blue line in \cref{fig:fvrel}, increasing the frequency of $50MHz$ from $2.85GHz$ to $2.90GHz$ at $0.9V$ will produce a $62.7mW$ power increase, while the same frequency increment, from $2.90GHz$ to $2.95GHz$ (involving an increase in the voltage level since $2.9GHz$ is the maximum allowed frequency for $0.9V$), produces a power increase of $740.7mW$, which is more than $10 \times$. The control algorithm should consider this difference when choosing the operating point of each \gls{pe}.

The example above is made more complex by the shared voltage among cores in the same domain, which introduces an additional coupling in the outputs of the controller (\cref{fig:component}). Increasing \reb{the} voltage \reb{consists in} increasing the power consumption (and temperature) of all the cores of the domain. Decreasing the voltage means capping the maximum frequency, and thus the performance attainable by the cores in the domain that were not hitting any power or thermal limit.
%the performance of the cores that could run at a frequency above the maximum allowed.

The control algorithm should consider this coupling when distributing the power or selecting the frequency of each core. This is well illustrated in the right-hand side of \cref{fig:fvrel}. For example, let's consider the case where core A is capped at $3.8W$ for thermal reasons, running at the operating point $\{2.75GHz, 0.85V\}$. %, where $2.75GHz$ is the maximum allowed frequency at $0.85V$. 
If, for external reasons (such as another core requesting a higher frequency), the domain's voltage is increased to $0.95V$, the core A's frequency would have to be reduced to $1.85GHz$ to retain the same power output. This will result in a performance loss due to the frequency being decreased.

Consider now the opposite case in which core A has thermal headroom and could run at its maximum speed. %, and the control algorithm has granted it the maximum power. 
If a different core B in the same voltage domain has to run at a lower voltage to dissipate some previously acquired heat\footnote{From the left-hand side of \cref{fig:fvrel}, a \gls{pe} running at the minimum frequency ($0.4GHz$) and at the maximum voltage, generates a considerable power consumption ($5.52W$). This is similar to the power generated by the same \gls{pe} at $\{3.0GHz, 0.95V\}$. This means that, even when running at minimum frequency, a core may need to reduce its voltage level to reduce its temperature.}, core A would be limited to the same maximum frequency as core B, and the power budget would not be used effectively (under-utilization).

Due to the coupling, a controller actuating a voltage increase should take into account whether the corresponding $\Delta P$ may incur power budget violations or high-temperature overshoots, requiring the thermal control to be activated to reduce the frequency (and thus performance). %, introducing also oscillations. 

\subsubsection{Disturbances}\label{sssec:ctrl_challenges:noise}
There are three different types of disturbance in a typical \gls{cpu} control scenario: instruction variability, process variation and drift, and sensor noise. The common control strategies available in the literature are able to tackle the magnitude of sensor noise, but such disturbance prevents precise power/workload identification at higher frequency domains.

Parameters deviation in modern and old technology nodes are related to the manufacturing flow, especially the front-end-of-line (FEOL). Previous works report that they can cause variations of up to $10\%$ in the current flowing through the manufactured devices~\cite{CMOS90_PVT_1, CMOS90_PVT_2, FINFET_PVT}.
Parameters can also slowly change during utilization due to aging. These deviations impact the power consumption and thermal response of the \gls{pe}. Control algorithms should take into account this differentiation and not treat each \gls{pe} equally.

Instructions variability, even if smoothed by the power delivery network, is a constantly varying noise input to the controlled system. This disturbance is unknown, varying with a frequency faster than the control algorithm's one, and accounts for a significant part of the power consumption $P(t)$ as shown in \cref{fig:leakage}.
The control algorithm can see power consumption fluctuation of several watts at each iteration with no changes to the input values of $F(t_k)$ and $V(t_k)$. 
Therefore the control action should provide a good balance between reactivity and filtering.

\subsubsection{Exponential relation of leakage power and temperature} \label{sssec:ctrl_challenges:exp}
The exponential relation between the leakage power and the temperature links the two objectives of the control problem (the power and the temperature) to each other. This relation amplifies the non-linearity already lying in the power map \eqref{eq:act_pow_map}, creates positive feedback, and is the main cause of the thermal runaway scenario described in \cref{ssec:ctrl_formulation}.  
Furthermore, this coupling hinders achieving a steady state, considering the disturbances present in the system that can continuously produce power spikes and oscillations. \reb{It also further complicates the challenges and coupling effects described in \cref{sssec:ctrl_challenges:var_coupling} - Controller outputs.}

\subsubsection{Power and thermal oscillations}\label{sssec:ctrl_challenges:osc}
Non-linearities, output discretization, and output coupling produce high-amplitude power consumption steps, that can cause significant oscillations around the thermal and power capping set points. For this reason, these set-points often cannot exactly be met. % by, for example, set-point tracking algorithms such as the \gls{pid}. %
These effects are exacerbated when dealing with an increasing amount of cores in the same shared voltage domain. Furthermore, in the latter scenario, the thermal coupling of neighboring cores worsen the oscillations.

%The presence of oscillations is caused by the control output discretization that makes harder to track exactly a setpoint, and the relation between the supplied voltage and the maximum frequency of the component, that introduce power output steps. The oscillations increased amplitude is caused by both the introduction of the variable voltage, which changes the power output of each component quadratically and with a \textcolor{red}{TODO:step}, and by the exponential relation between the leakage power and the temperature state of the core and the voltage of the component, that further increase the amplitude of the power step.

%Even if linked, thermal and power limits have different characteristics. As presented in the \ref{}, the temperature of a \gls{pe} depends on its temperature state, power consumption, and the temperature state of neighbour elements and it has time constants greater than $1ms$. Instead, the power consumption is instantaneous, and even if it depends on the temperature state of the \gls{pe}, it mainly depends on the controllable Frequency and Voltage and the type of executed workload which is unknown.

%The modifications in the system model described in section \ref{sec:background} introduce several new control challenges with respect to the control problem analyzed in \cite{CCTA}.

%As we show in \textcolor{red}{TODO}, also domains with higher amount of cores has higher presence of oscillations due to the fact that a voltage change will affect several cores in the region, which they are also thermically related to each other.

High-amplitude oscillations violate thermal and power capping constraints, decrease performances over time, and stress the hardware, causing it to age faster.

%This challenge is mostly present at high temperatures and voltages close to the $T_{\text{CRIT}}$, where the exponential relation has the higher \textcolor{red}{ratio}.

%For these reasons, the bigger amount of cores a domains has, more these complication will be severe, and more difficult the choice of the right voltage will be.

%\textcolor{red}{TODO: not here  but remember to talk about "so the soa decided to use max" fuxxy is good because it tends to reduce voltage.}

%AFTER our previously presented control algorithm which is b. To avoid treating with non-linear control, we chose to perform our control action on the power

%domains stuff
%frequency vs voltage vs performance
%freuqnecy non.linear vs power (model) vs knowing V in advance
%thermal runaway

%Present the Workload issue/noise

%%%%%% TODO:

\subsubsection{Controller delays}\label{sssec:ctrl_challenges:delay}
\reb{To be consistent in the output generation, digital discrete controllers such as the \gls{llc} apply the computed output at the beginning of the next periodic execution~\cite{discrete_book}, thus adding a delay of 1 $t_s$, with $t_s$ being the controller execution periodicity interval.

In the described control problem, there is an additional source of incremental delay given by the workload $\omega$. The workload directly affects the instantaneous power consumption of the \glspl{pe} \eqref{eq:act_pow_map}, but it will only be measured at the following controller interval. If the workload has a sudden change in its average value, which is a common event, the controller response will have a delay of $(1 + \upsilon)t_s$, where $\upsilon \in \: (0, 1)$ is the latency introduced by the workload variation.

This delay is an additional cause of power and thermal oscillations. To reduce the thermal oscillations, a controller interval $t_s$ less than at least half the faster thermal time constant is required.

To reduce power oscillations caused by workload variation instead requires \gls{pe}'s hardware power management features added by the manufacturer. In the past, Intel was dealing with this problem by forcibly restricting the frequency when certain high-power workloads were detected~\cite{schone_energy_2019}, new designs include microarchitecture mechanisms to delay high-power instruction dispatching, spreading the power increment into multiple $t_s$~\cite{neoverse_v2}.
}

\section{Control Algorithms}\label{sec:methods}

In \cref{ssec:oca_extended}, an extended and improved version of the control algorithm proposed by Bambini~\textit{et al.} in~\cite{ICCD_bambini} and evaluated in~\cite{CCTA_bambini} is presented to account for the not-idealities in the control problem introduced in this work. The original version is referred to as \textit{\gls{ba}} and the enhanced version as \textit{\gls{oca}}. A comparison between the performance of the BA and \gls{oca} against the \textit{industrial \gls{sota}} is then carried out. As described in the related work section, this is based on a \gls{vba}. Results will show that the \gls{oca} underperforms the industrial \gls{sota} when the not-idealities are present. In \cref{ssec:fuzzy} a \gls{fca}, the proposed algorithm designed to overcome the limitations of the \gls{oca}, is described. 

%In the following section, we recall the \gls{oca} presented in Bambini~\textit{et al.}~\cite{ICCD_bambini} and propose three modifications to account for the introduced output constraints and couplings. Then, we detail the proposed algorithm (FCA), which accounts for several of the control challenges presented in \cref{ssec:ctrl_challenges}.
%as well as the main \gls{vba} employed by most of industry vendors, such as IBM.
%Last, in \cref{ssec:methodology_fig_merits} we introduce the evaluation methodology and figure of merits used for the experimental assessment carried out in \cref{sec:results}.

\subsection{Addressing the non-idealities in the original algorithm}\label{ssec:oca_extended}

\begin{figure*}[t]
	\centering
	\includegraphics[width=1\linewidth]{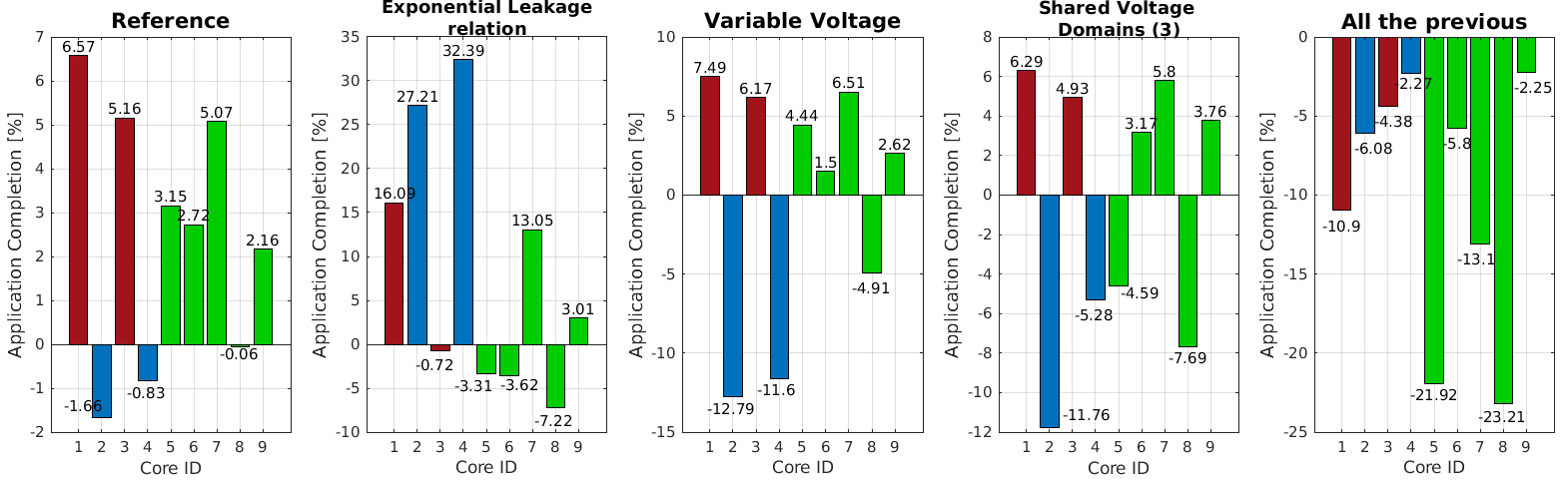}
	\caption{\label{fig:old_results} Five iterations of the test presented in Bambini et al. \cite{CCTA_bambini} with the additional thermal and power model features introduced in \cref{ssec:addthermmodel}. From the left-hand side, (1) the original test, (2) the addition of the exponential leakage relation \eqref{ssec:expleak}, (3) the addition of \gls{dvs}, (4) the addition of 3 shared voltage domains, and (5) all the previous together.}
\end{figure*}

The BA has the following key elements: (i) an initial conversion step (\textit{Conv2P}) in which $F_T$ and $\omega$ are used to compute an estimation of each core's power consumption $P_{\text{est}}$, (ii) a cascade of two blocks to perform the power capping and the thermal capping respectively, that outputs the reduced power $P_a$, and (iii) a final conversion step (\textit{Conv2F}) to compute from $P_a$ the frequency setting for the core $F_a$.

\Cref{fig:old_results} shows the comparisons of BA and \gls{oca} against the \gls{vba} in a simulation of a computing tile of $9$ \glspl{pe} simulating a workload lasting $2s$. Each of the subplots in \cref{fig:old_results} reports on the y-axis the relative speedup of each \gls{pe} (reported on the x-axis). Different colors are associated with different types of workloads: high-power workloads are red (core 1, 3), low-power are blue (core 2, 4), and variable workloads are green (core 5-9). \reb{\Glspl{pe} target frequency is set to the maximum to stress the control algorithm in the worst capping scenario.} Each subplot introduces in the simulation environment one of the not-idealities described in \cref{sec:background}. At the same time, we add a different control algorithm feature implemented in the \gls{oca}. %\Cref{fig:old_results}.1 introduces in the plant model the newly proposed thermal models as described in \cref{sec:background}.  \Cref{fig:old_results}.2 introduces in the plant model the exponential leakage model described in \cref{ssec:expleak}. \Cref{fig:old_results}.3 introduces in the plant model the shared voltage domains described in section bla bla and the \texit{enhanced baseline} controller.  \Cref{fig:old_results}.4 combines in the plant models all the previously introduced not-idealities and the \texit{enhanced baseline} controller. 

\reb{\Cref{fig:old_results}.1 shows that the more accurate thermal model introduced in this work with the addition of the secondary thermal path and other thermal dissipation elements leads to similar results to the ones obtained in~\cite{CCTA_bambini}.}

%The \gls{oca} yielded better results compared to \gls{sota} with the assumption of fixed voltage, no shared-voltage domains, a relaxed step-linear relation between power and temperature, and a fine-grained Frequency discretization~\cite{CCTA_bambini, SAMOS}.

%\Cref{fig:old_results} shows the performance of \gls{oca} with several thermal and power model configurations, one for each modeling contribution of this work. To assess the quality of the control, we measure the percentage of execution completion of a simulated workload within $2s$ on a computing tile of $9$ \glspl{pe}. The results are compared to a reference design. Higher values correspond to faster execution. %
%\Cref{fig:old_results}.1 represents \gls{oca} with the thermal model described in \cref{ssec:addthermmodel}. The results recall the ones obtained in~\cite{CCTA_bambini}. %
%
In \cref{fig:old_results}.2, the exponential relation between leakage power and temperature is introduced. With a fixed high voltage, the power consumption $P_i(t)$ and the temperature $T_i(t)$ grow faster, affecting the control quality. %
In \cref{fig:old_results}.3 \gls{dvs} is enabled, and the \gls{oca} is switched in place of the \gls{ba} in the comparison with the \gls{vba}.
As presented in \cref{ssec:ctrl_challenges}, different voltage levels have severe variations in the power consumption (and temperatures), more than frequencies. In BA, the voltage has to be chosen in advance since the first step of the algorithm involves the conversion of the input variables into $P_{\text{est},i}$ (Conv2P). 
In the scenario of a voltage domain per each core, this issue could be solved in Conv2F by selecting a pair of $F_i$ and $V_i$ according to a look-up table. 
Nevertheless, due to the non-idealities of the actuator described in \cref{ssec:realact}, such implementation incurs an over-conservative policy in the presence of shared voltage domains. Different cores may have their power reduced for different reasons, and the choice of a voltage \emph{a-posteriori} could severely affect the performance objectives. %
As shown in section (\ref{ssec:ctrl_challenges}), increasing the voltage of the domain can cause a performance drop to all the other cores of the domain. A better power distribution would balance the performance of all the cores according to the final voltage choice.

To address such over-conservative behavior and prevent performance losses due to ineffective \gls{dvs}, a moving average filter for the voltage selection is introduced in the \gls{oca}. %
After each iteration $t_{k}$ of the control algorithm, the difference between the target frequency $F_T(t_{k})$ and the applied frequency $F_a(t_{k})$ is added to an accumulation value $F_{\text{MA}}(t_{k})$ through a forgetting factor $\lambda_{\text{MA}}$. With several iterations, $F_{V-MA} = F_{T} - F_{MA}$ should be tracking the average frequency chosen by the control algorithm over a period $\mathcal{T}_{\text{MA}}$ dependent on the choice of $\lambda_{\text{MA}}$. In the Conv2P part, $F_{\text{MA}}(t_{k})$ is used to select the initial shared voltage of the domain to have a more realistic evaluation of the power output of each component. However, the target frequency $F_T(t_{k})$ is still used in the computation of the estimated power $P_{\text{est}}$ to allow the control algorithm to increase the applied frequency when thermal and power headrooms are available. %
At each iteration, the \gls{oca} executes:
\begin{align}
    F_{V-\text{MA}}(t_{k}) &= F_{T}(t_{k}) - F_{\text{MA}}(t_{k-1}) \\
    V(t_k) &= f_V \big(  F_{V-\text{MA}}(t_{k})\big) \\
    P_{\text{est}} &= h_{\text{est}} \big( F_{T}(t_{k}), \, V(t_k), \, \cdots \big) \\
    F_{\text{MA}}(t_{k}) &= (1 - \lambda_{\text{MA}})F_{\text{MA}}(t_{k} - 1) \; + \\ & \lambda_{\text{MA}}(F_{T}(t_{k}) - F_a(t_{k}) ) \label{eq:MA}
\end{align}
where $f_V(F_i)$ is a step-wise monotonically increasing function that, given a frequency $F_i$, computes the minimum voltage such that \eqref{eq:FV} is fulfilled.

\reb{\Cref{eq:MA} can be thought of as a discretized version of a low-pass filter where the forgetting factor $\lambda_{\text{MA}}$ is equal to the ratio between the controller sampling time $T_s$ and filter time constant $\tau$.} \reb{The value of $\tau$ should be chosen to reject small system state perturbation caused by workload variation and intercept the main tendencies. In this work the value of $\tau$ is chosen as $\tau = 1.25\cdot 10^{-2}$.}
This method decreases the performance during transients after a change in one of the given set points or system conditions, depending on the choice of the forgetting factor. A strategy to detect \reb{application phases} could be employed to partially avoid this performance reduction, by resetting $F_{\text{MA}}(t_{k})$ after a \reb{ different phase is detected or a set point change is required from \glspl{hlc}}, significantly decreasing the ``adaptation time''.
Observing \cref{fig:old_results}.4, the addition of shared voltage domains and a variable voltage lead to results similar to the original test from~\cite{CCTA_bambini} shown in \cref{fig:old_results}.1.

A second necessary modification to the BA is due to the larger difference between the measured and computed power $P_{\text{est}}$, originating from the exponential relation highlighted in \eqref{eq:exp_leakage}. This can lead the BA to miss the power budget set point by a large and variable margin (positive or negative). This approach is in contrast to classical predictive approaches based on an a priori knowledge of the power model \cite{TPDS}, but this would require a non-negligible effort to study all different operative points. %
Our solution is simpler and employs a moving average adaptation filter. The difference between the sum of $P_{\text{est}}$ and the measured power is added in the power distribution block of the algorithm to take into account the power model difference through a moving average filter to avoid high-frequency noise and oscillations.

A third observation is how to choose the voltage that is shared among all the cores in the domain. In \gls{oca}, each core requests a voltage level based on the minimum voltage required to run at the given frequency (in this case, the $F_{V-\text{MA}}$). \Gls{sota} in industry chooses the maximum voltage, i.e., the voltage level that is compatible with the highest frequency requested in the voltage domain, according to \eqref{eq:FV}. This approach causes a larger performance drop if frequencies in the same domain have larger value differences. This behavior is shown in \cref{fig:old_results}.5 where the \gls{oca}'s results are worse than \gls{sota}.

Other control decisions to be investigated are (i) the selection of a voltage as a percentile of the casted vote (e.g. $90\%$), instead of the maximum, or (ii) running an optimization step that computes the voltage that will cause the minimum sum of performance differences according to an `importance matrix' $\mathcal{R}$.%\footnote{This is assuming that all the cores have the same 'importance'. A previous step could be to identify cores that are idling or memory-bound and remove those from the sum.}.

We can conclude that the \gls{oca} approach is too simplistic to cope with the power and thermal management of many-cores systems when non-idalities are considered. In the next section, we propose a different algorithm to tackle them.

\subsection{Proposed Fuzzy-inspired Iterative algorithm}\label{ssec:fuzzy}

%Presented above, are some adjustments to the original control algorithm to face the new control challenges introduced by the non-idealities in the updated system model. 
As from \cref{fig:old_results}.5, there are some scenarios where the \gls{oca} does not perform as well as the reference. To further improve control quality, the control algorithm was redesigned retaining its advantages while addressing at the same time some of the criticality observed in the previous section.

The algorithm is still based on a cascade design with power as the main control variable. The problem with the \emph{a priori} choice of the shared voltage in the domain is solved by employing an iterative solver method in the Conv2F part, converting the power into the pair of frequency and voltage.

The \gls{sota} approach of selecting the maximum voltage can be expressed as $ V_{\mathcal{D}_j} = \max \bigl( f_V (F_{i \in \mathcal{D}_j}) \bigr) $. 
Since $f_V(F_i)$ is step-wise monotonically increasing, the right term can be rewritten as:
\begin{equation} \label{eq:indstand}
 V_{\mathcal{D}_j} = f_V\bigl(\max(F_{i \in \mathcal{D}_j})\bigr)
\end{equation}
Denoting $h_{\text{est}}$ as mapping \eqref{eq:pow_map} of the control algorithm to compute $P_{\text{est}}$ and $P_a$ as the output of the control algorithm, a system of equations for each domain $\mathcal{D}_j$ can be derived where $P_a$ equals $h_{\text{est}}$ combined with \eqref{eq:indstand}.
This way, the frequency-voltage relation \eqref{eq:FV} is directly enforced in the conversion step. 
\begin{equation} \label{eq:covers}
	 \begin{cases}
	    {P}_{a,1} = h_{\text{est}} \big( F_1, \, f_V\bigl(\max(F_{i \in \mathcal{D}_j})\bigr), \, \cdots \big) \\
	    \quad \vdots \\
	    {P}_{a,n_j} =  h_{\text{est}} \big( F_{n_j}, \, f_V\bigl(\max(F_{i \in \mathcal{D}_j})\bigr), \, \cdots \big)
	\end{cases}
 \\[15pt]
\end{equation}
%where on the right side of the equation is the approximated power model.

Assuming $ \forall \; i = 1, \ldots, n_j$
\begin{equation} \label{eq:limits}
\begin{aligned}
   {P}_{a,i} &> \min \bigl[  h_{\text{est}} ( F_{\text{min}}^S, \, V_{\mathcal{D}_j}, \, \cdots ) \bigr] \\
    {P}_{a,i} &< \max \bigl[  h_{\text{est}} ( F_{\text{max}}^S, \, V_{\mathcal{D}_j}, \, \cdots ) \bigr]  
    \\
\end{aligned}
%\qqad \forall \; i = 1, \ldots, n_j
\end{equation}
a solution to the system of equations \eqref{eq:covers} is guaranteed to exist.

%The conversion step thus become a root-finding problem of a system of non-linear equations for each voltage rail in the system:

To solve these $j$ non-linear systems, the global Newton-Raphson iterative method could be employed, using the Softmax function $\mathcal{S}_{\alpha }$~\cite{softmax_prop} as an approximated $\max()$ function that is differentiable. Even with hardware-optimized algorithms, the time to solve the global N-R with a reasonable tolerance is not feasible for a real-time control solution running on an embedded \gls{llc} with tight periodicities such as~\cite{Ottaviano2023ControlPULPAR}.

Considering that the solutions' accuracy is bounded by the precision of the system actuators (generally not smaller than $5\cdot10^{-3}$), the bisection method~\cite{burden2015numerical} was chosen instead. The bisection takes more iterations to converge, but each iteration is faster compared to Newton-Raphson. \reb{Furthermore, it does not require inverting the Jacobian and its execution time is less dependent on domain dimension}. From a deployment perspective on an embedded \gls{llc}, this means reducing the execution time, hence being able to execute within the control period. 

Since the bisection does not need derivatives, the $\max()$ function can be employed. Furthermore, the function $f_V$ can be upper-bounded by a non-linear polynomial function, such as the second order function $f_V(F_i) = F_i^2 + k_1F_i + k_0$. The coefficients of this function can be found \emph{a priori} since they depend on the F-V relation \eqref{eq:FV}.
This solution still didn't address the problem of the output discretization as both the F and V values are considered continuous. %For this reason a discretization step is taken afterwards.

\reb{However, using an iterative method for the Conv2F part instead of the $F_{\text{MA}}$ in conjunction with the original \gls{pid} control algorithm, yielded poor results.
The \gls{pid} is a linear time-invariant controller. Albeit the thermal evolution of the system can be described by a linear relation, as non-idealities and coupling such as the ones described in \cref{ssec:ctrl_challenges} are introduced in the model the limitations of the \gls{pid} emerge.
In particular, the discrete coupled outputs which are exponentially related to the temperature state, require a \gls{pid} tuning pushed towards a fast response to avoid the thermal runaway scenario. These types of \gls{pid} tuning also generate high-frequency output oscillations which cause control performance losses, and these oscillations were more frequently present in combination with the iterative method.}
%In particular, choosing the operating point directly in combination with the \gls{pid} thermal capping was causing high-frequency output oscillations.
%as it was not working well with the thermal PID capping component of the control algorithm. In particular, having the operating point being chosen instantly would cause oscillations \textcolor{red}{write this: praticamente se il punto operativo è scelto instantaneamnete invece che un po' più dolcemente con il MA questo causava forti oscillazioni nell'output perché il PID cercava di traccare il punto di massima temperatura ma l'output variava tantissimo da una iterazione all'altra.}
 %
For this reason, the thermal capping control was changed to a heuristic LUT-like control inspired by the Fuzzy control theory. %

The fuzzy control theory is a simple and effective way of turning symbolic decision tables into control laws. In practice, fuzzy logic consists of three main steps: (1) the control inputs are converted into fuzzy quantities (i.e., sets) through a process called \emph{fuzzification}; (2) logic rules are applied to decide the control action, and (3) the latter is finally converted into a quantitative control output value in a process called \emph{defuzzification}~\cite{nguyen2012fuzzy}.
%a set of rules can be ``compiled'' into a look-up table or a mathematical function, expressing the control value in terms of the inputs of the controller. After the control law is applied, the output is then returned into values used by the actuators in a process that is called ``defuzzification''  \textcolor{red}{TODO: check christian}
%
Fuzzy control can be applied to non-linear systems, it does not need a model, it is simple to understand and compute, and it is modular and easily modifiable. %
Hereafter, we refer to the fuzzy-inspired control with the iterative operating point selection as \gls{fca}.

Due to the capability of fuzzy logic to directly tackle non-linearities, the thermal control can be directly applied to the frequency $F$ and voltage $V$ values. For this reason, in \gls{fca} the thermal capping block is moved before the power distribution block. With this structural modification, already distributed power from the power capping block is not wasted to a later potential further power reduction for thermal capping reasons. 
The control structure in \gls{ba} was motivated by the power affecting the temperature, and our proposed control mimicked that structure. But in the updated model, the temperature is affecting the power too, making the previous statement no longer valid.
%\textcolor{red}{TODO: show new picture with the structure}

In the \gls{fca}, a \emph{fuzzy state} representing the reduction in frequency and voltage due to thermal capping reasons is assigned to each \gls{pe}. \reb{A \gls{lut} is created by dividing \glspl{pe}} current temperature $T_i(t_k)$ and its derivative $\Delta{T}_i(t_k)$ into regions relevant to the thermal control, and is populated with values to either increase or decrease the fuzzy state. The derivative is approximated as the difference between the current temperature and the temperature of the previous $s$-th step $\Delta{T}(t_k) = T(t_k) - T(t_k-s)$, as a means to filter the sensors' noise. The choice of $s$ depends on the ratio between the fastest thermal time constant of the system and the execution period of the thermal part of the control algorithm.

\begin{table*}[t]
\centering
\begin{tabular}{r|c c c c c} 

\textbf{$\Delta$T/T[\degree C]} & $\;\; \mathbf{<45} \;\;$ & $\mathbf{45-65}$ & $\mathbf{65-80}$ & $\mathbf{80-T_L}$ & $\;\; \mathbf{\geq T_L}\;$  \\ 

\hline
$\mathbf{<0}$                             & 2  & 2          & 1          & 0              & -1         \\ 
\hline
$\mathbf{0-0.5}$                           & 2  & 1          & 1          & 0              & -1         \\ 
\hline
$\mathbf{0.5- 1.0}$                           & 1  & 1          & 0          & -1             & -2         \\ 
\hline
$\mathbf{1.0- 2.0}$                          & 1  & 0          & -1         & -2             & -3         \\ 
\hline
$\mathbf{\geq 2.0}$                           & 0  & 0         & -2         & -3             & -4         \\
%\hline
\end{tabular}
\caption{\label{tab:fuzzy} LUT table to determine the increment/decrement of the fuzzy state on each iteration based on the derivative of the temperature $\Delta{T}(t_k)$ and the temperature value $T(t_k)$. }
\end{table*}

%
%Due to the analysis shown above in the different contribution of F and V to the output power \textcolor{red}{TODO: rewrite with names}, we choose to favor the reduction of V
Each negative integer value of the fuzzy state decreases half of the frequency step between two consecutive voltage levels, meaning that eacheven negative fuzzy state value reduces one voltage level. %
%
%With this configuration, differently from the \gls{pid}, the thermal control starts reducing the operating point before reaching the thermal limit, based on the derivative while directly operating on the frequency and voltage values. %
%
This configuration could also incorporate a frequency turbo boost feature through positive fuzzy states. Since we do not analyze the turbo capabilities in this work, the fuzzy states are saturated to 0 to maintain a balanced comparison with the other algorithms.

\reb{The choice of the number of columns and rows of the \gls{lut} depends primarily on the memory and computation constraints of the \gls{llc}, and secondary on the desired quality of the thermal control. The choice of the thresholds values of the derivate mainly depends on the thermal time constants of the system, while the choice of the temperature thresholds mainly depends on the slope of the P-T leakage relation (\cref{fig:leakage}). The \gls{lut} values can be chosen by discretizing a surface describing a trade-off between performance reduction and control performance. The \gls{lut} used in this work is showed in \cref{tab:fuzzy}.
}

In \gls{fca} the power distribution part is also slightly modified to reflect the presence of voltage-sharing domain configurations. The BA power distribution algorithm is applied to each domain rather than each core. A linear combination between the maximum temperature and the average temperature of the cores in each domain along with the sum of all domain's cores $P_{\text{est}}$ are used as inputs of the heuristic power capping algorithm. Then, for each domain, the power is distributed based on the workload $\omega_i$ of each core, trying to favor high-demanding cores.

\section{Evaluation methodology}\label{sec:methodology_fig_merits}

% We do some kind of taxonomy and sota description in the related works. Either we add 'comprehensive' (like in a survey) or we remove the sentence, imho.
%A categorization of the state-of-the-art thermal and power control designs and algorithms, which solve control problems similar to \eqref{eq:purp} is out of the scope of this work. 

A fair comparison among control algorithms is challenging to achieve, for multiple reasons. 
First, the control quality is tightly coupled with the microarchitecture of the system to be controlled (i.e. the \gls{hpc} processor). Second, there is no standardized metric for this comparison, since software benchmarks compare the overall \gls{hpc} processor performance and are not focused on the controller alone. Finally, it is difficult to find updated information, algorithms, hardware descriptions, etc. that are covered behind industry secrets.

\reb{The most popular industry-standard control is the Voting Box Algorithm (VBA)~\cite{rapl_2015, AMD, ibm_occ}, where several operating points are computed independently one from the other, and then the minimum is selected as the output of the controller.}
One of the available resources for comparison is the IBM OCC control algorithm from the IBM Power9 chip, whose software code is fully open-source \cite{ibm_repo}. We use this baseline for a reproducible comparison.

\reb{Including in the comparison the \glspl{hlc} presented in the Related Work would lead to misleading results. As described in \cref{sec:related_works}, \glspl{hlc} and \glspl{llc} are designed to work together and they are developed for their specific objectives, execution scope, and time domain. Despite being able to leverage higher computational capabilities, \glspl{hlc} would be outperformed by \glspl{llc} in the comparison due to their higher controller step time, their assumptions and abstractions of the system (in particular, the assumption that an \gls{llc} beneath will be controlling higher dynamics), and the absence of significant features, such as receiving an operating point or capping the power.

For these reasons, the test will include a comparison between} the reference \gls{vba} with both the \gls{oca} discussed in \cref{ssec:oca_extended}, and the \gls{fca} introduced in \cref{ssec:fuzzy}, to observe the benefits and the drawbacks of each design. %

We identify three different workload scenarios to investigate the response of each control algorithm: (i) the MAX-WL test, where all \glspl{pe} execute a vector workload consuming the maximum power and the target frequency is set to the maximum frequency, (ii) the MULTI-WL test, where some \glspl{pe} execute vector workload at the maximum frequency ($3.45GHz$), others perform a mix of floating point and int workload at a lower frequency ($2.7GHz$), while the rest of the cores (the majority) idle at the minimum frequency ($0.4GHz$); each voltage domain includes at least one \gls{pe} from both the first two groups to stress the voltage coupling control constraint, and (iii) the CLOUD-WL test, where random and varying workload is executed on all cores, while the maximum frequency is always required. %
To see the control algorithm fully in action, the power budget of the processor is changed four times during each test, with both increasing and decreasing patterns.

These tests are evaluated on three different variants of the thermal model \eqref{eq:lumped_par_mod_comp} to stress the adaptability of each control algorithm to different cooling conditions. We identify a water-cooled model (WATER) where the temperatures of the cores are more homogeneous, an air-cooled model (AIR) where the temperature of the cores has a Gaussian-shaped spatial distribution, and a rack air-cooled model (RACK) where air is pushed from the front and exhausted on the back, generating a column-wise spatial distribution of the temperatures modeled with an additional column-wise temperature coupling among cores.

%\textcolor{red}{Add and describe pictures}:

These tests are also evaluated on four different voltage domain configurations to analyze the impact of voltage coupling and the adaptability of each control algorithm. The configurations are: (i) all cores share a single voltage domain (1-D), (ii) the cores are equally divided into 4 domains (4-D), (iii) the cores are equally divided into 9 domains (9-D), and (iv) each core has its own voltage (A-D), to simulate the case without voltage coupling among cores.

The main metrics to be evaluated when comparing tests' results can be directly derived from the control objective \eqref{eq:purp}. These can be translated to:
\begin{itemize}
    \item \underline{Thermal capping}:
        \begin{itemize}
            \item Exceeded value: maximum exceeded temperature [\degree C] compared to the capping target
            \item Total time: percentage of time [\%] exceeding the capping target.
        \end{itemize}
        \reb{Since the thermal threshold is taken below the critical thermal limit (\ref{ssec:ctrl_formulation}), what would cause the system to fail is the maximum exceeded temperature. For aging and degradation, the metric to consider is the amount of time that the threshold is exceeded.}
    \item \underline{Power capping}: 
        \begin{itemize}
            \item Exceeded average value: the ratio between the average exceeded power with its power budget target [\%] 
            \item Total time: percentage of time [\%] exceeding the capping target\footnote{We consider only the time when such excess is possible. For example, when there is no power budget, it cannot be exceeded.}.
        \end{itemize}
    \reb{It is important to know the average exceeded power and time since spikes are tolerated and smoothed by hardware features~\cite{PDN}.}    
    \item \reb{\underline{Target Compliance}}:
    \begin{itemize}
        \item The L2-norm\footnote{Assuming $\mathcal{R}$ in \eqref{eq:purp} as an identity matrix, meaning all the cores have the same importance.} of the difference $\Delta F_{pe} = F_T - F_a$, normalized by the number of cores and control steps [$GHz/s$];
    \end{itemize}
\end{itemize}

Other figures for the comparison are \reb{the \textbf{average} and \textbf{min execution progress (AV/MIN-$Wl_p$)} across all cores, as the amount of workload executed [\%] with respect to the case where there is no control action. This serves as a direct index for the impact of the control algorithm on the application execution, with the MIN-$Wl_p$ metric describing the most penalized core.}

%\begin{itemize}
    %\item \underline{\gls{hpc}-related}:
    %\begin{itemize}
        %\item Average Temperature [\degree C]
        %\item Average and min Performance (AV/MIN-$Wl_p$) across all the cores, as amount of workload executed %\footnote{In some tests, the absolute value of the $Wl_p$'s metrics will be identical to the $\Delta F_{pe}$'s ones, but in others, due to the fact that memory-bound portion of code are simulated where cores' frequency will not impact execution speed, they may distance themselves.}, 
        %[\%] with respect to the case where there is no control action
    %\end{itemize}
   % \item \underline{Control-challenges-related [\ref{control challenges}]}:
    %\begin{itemize}
    %    \item Frequency oscillations. These are computed as the standard deviation among windows of $1ms$, $10ms$, $100ms$, and $1s$. \textcolor{red}{TODO: Finish describing: we will give the average, sd of average, and mean of sds.}
        %\item Utilization of the power budget
    %\end{itemize}
%\end{itemize}

\begin{figure*}[t]
	\centering
	\includegraphics[width=1\linewidth]{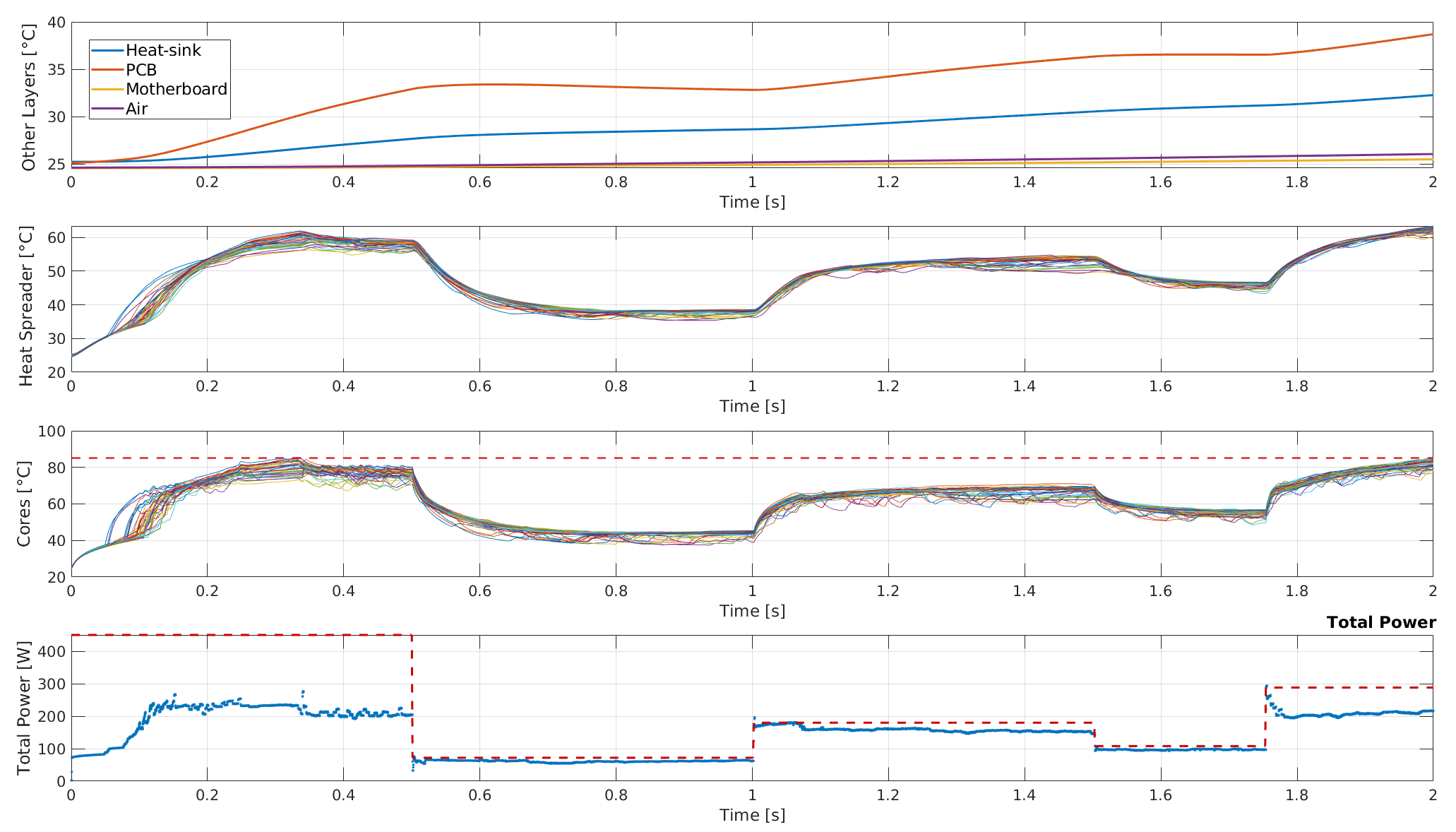}
	\caption{\label{fig:show} The thermal and power evolution of the proposed comparison test. In particular, this is an AIR model with a 4-D configuration, executing the CLOUD-WL and it is controlled by the \gls{fca} algorithm. \reb{The first three rows show the temperature evolution of the thermal dissipation paths (1), the heat-spreader hotspot blocks (2), and the cores (3). The last row shows the total consumed power measured. The red dashed lines represent the thermal and power limits.}}
	%\vspace{-15pt}
\end{figure*}

The power and thermal model have a randomly generated (but constant among all tests) set of parameter uncertainties, and all temperature sensors have a white noise with an amplitude of $1 \degree \text{C}$ as it is commonly found on \gls{pvt} sensor' specifications~\cite{TCAS}.
The tests run for $2s$ using a \gls{mil} framework consisting of an emulated computing chiplet with 36 cores. \reb{All the tests are executed for 10 iterations with different initial conditions, to render the results independent from the initial state of the system.}
An example of the test is shown in \cref{fig:show} as a reference for the power and thermal evolution.

\section{Experimental Results}\label{sec:results}

\reb{Results data are aggregated into figures according to workload type (\ref{fig:wl}), Domain number (\ref{fig:dom}), Model type (\ref{fig:mod}), and tests iteration (\ref{fig:test}). An additional figure with violin plots (\ref{fig:violinplot}) serves to describe the result distribution among all tests.}

\subsection{Simulation Results}

\begin{figure*}[t]
	\centering
	\includegraphics[width=1\linewidth]{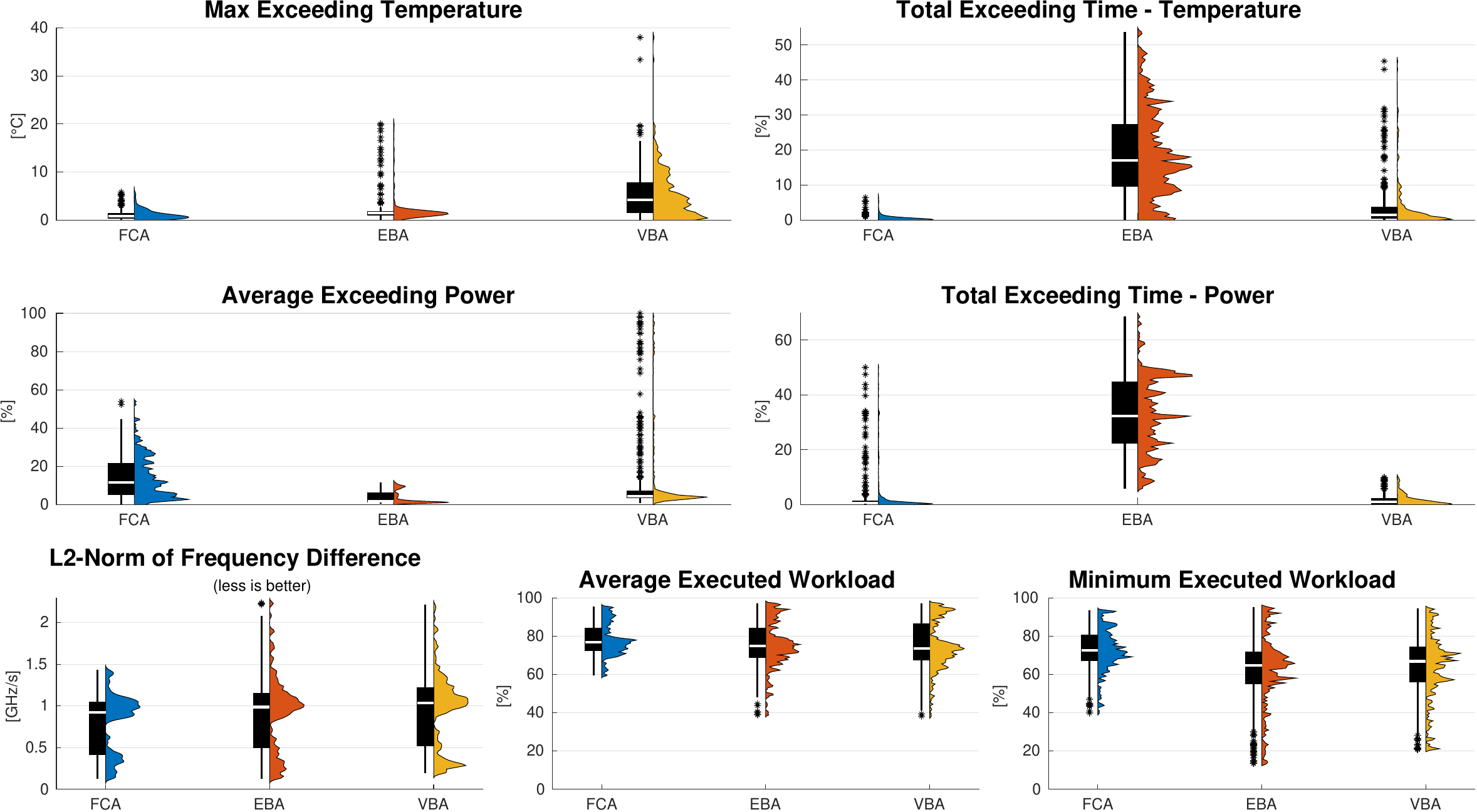}
	\caption{\label{fig:violinplot} \reb{Overall distribution of the control performance metrics plotted per algorithm. In the first row are the Thermal metrics, in the middle row are the Power metrics, and in the last row are the Target Compliance and execution progress metrics. }}
	%\vspace{-15pt}
\end{figure*}

\subsubsection{Thermal Capping Metrics}
\reb{
\Gls{fca} outperforms both \gls{oca} and \gls{vba} in both thermal capping metrics, and has the least variation across tests, suggesting that the proposed fuzzy-inspired thermal design is an improvement over the other \gls{pid} controllers. 
It achieves a maximum exceeded temperature that is $2.44$ and $5.52$ times lower than \gls{oca} and \gls{vba} respectively, while having more than $90\%$ lower exceeding time. 

\Gls{oca} has the highest average exceeded time ($18.82\%$), but with a low maximum exceeded temperature across tests ($1.94\degree \text{C}$ on average). This means that a lower thermal margin of about $2 \degree \text{C}$ could be defined to remove almost all the exceeded time.} Furthermore, \gls{oca} performs better as the number of domains increases. This is probably a heritage of the algorithm being developed with no voltage coupling constraints.

\Gls{vba} has the worst thermal capping performance among the three analyzed algorithms. 
It achieves the highest maximum exceeded temperatures (which are dangerous for the hardware) and struggles particularly in the RACK model and CLOUD-WL tests where it reaches an exceeded maximum temperature of \reb{almost $20 \degree \text{C}$ with two spurious results over $30 \degree \text{C}$. The larger the voltage domains are, the worse it performs, indicating difficulties in managing voltage couplings.} 
%The average temperatures are similar to \gls{fca} and generally below \gls{oca}, meaning that the high maximum exceeded temperatures are a transient problem and not a steady-state issue.

%Analaysis per domain change??

% analysis per model change?

%analysis per wl change?

%The MULTI-WL test is the one in which the difference in thermal capping performance with the other two algorithms is the minimum. The MULTI-WL test is the one in which the VBA will perform better across all the different metrics.

\begin{figure*}[t]
	\centering
	\includegraphics[width=1\linewidth]{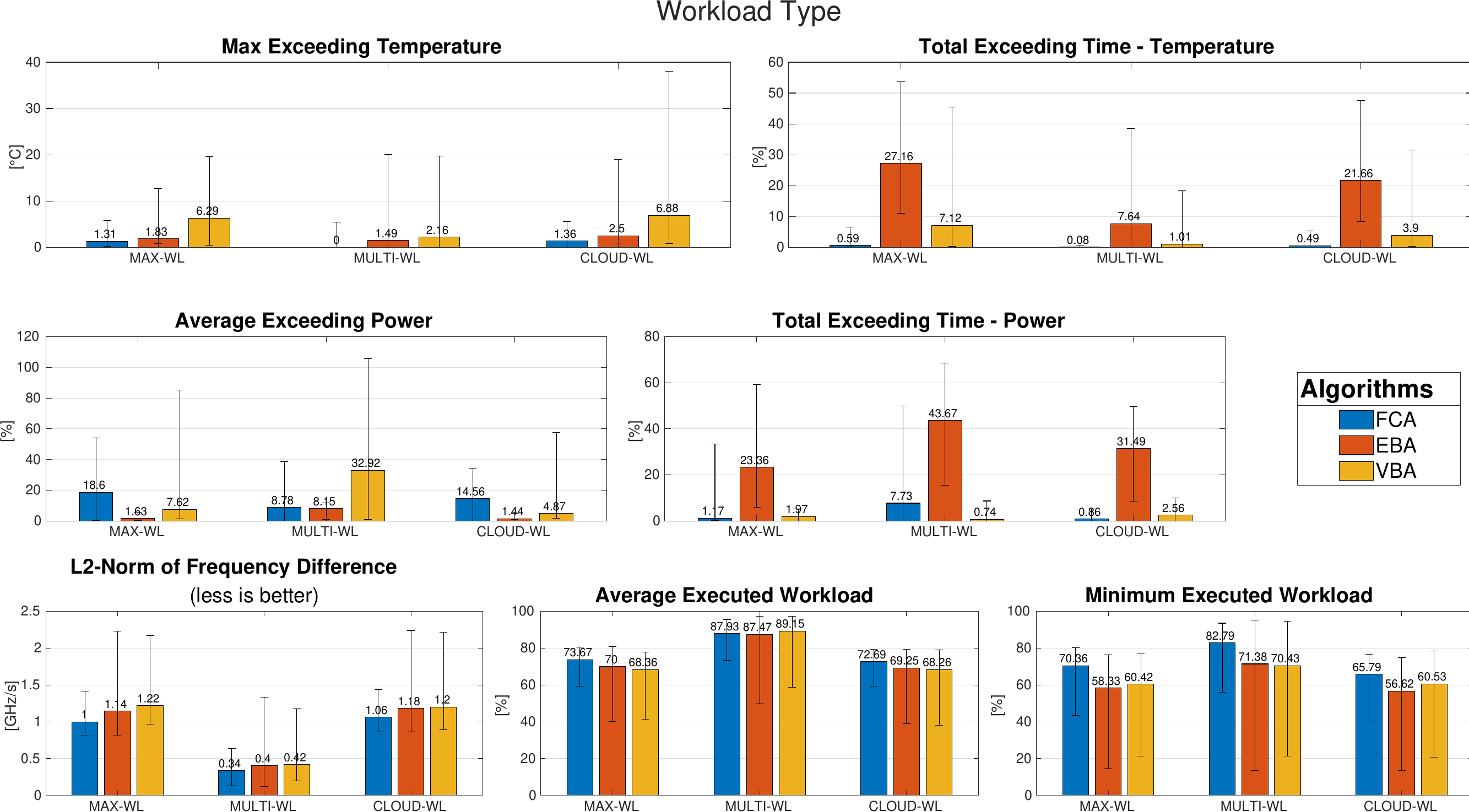}
	\caption{\label{fig:wl} Control performance metrics plotted per workload type (vector -MAX-WL-, combination of vector, INT/Float, and Idle -MULTI-WL-, and random varying workload -CLOUD-WL-) with their relative error bars. In the first row are the Thermal metrics, in the middle row are the Power metrics, and in the last row are the Target Compliance and execution progress metrics.}
	%\vspace{-15pt}
\end{figure*}

\begin{figure*}[t]
	\centering
	\includegraphics[width=1\linewidth]{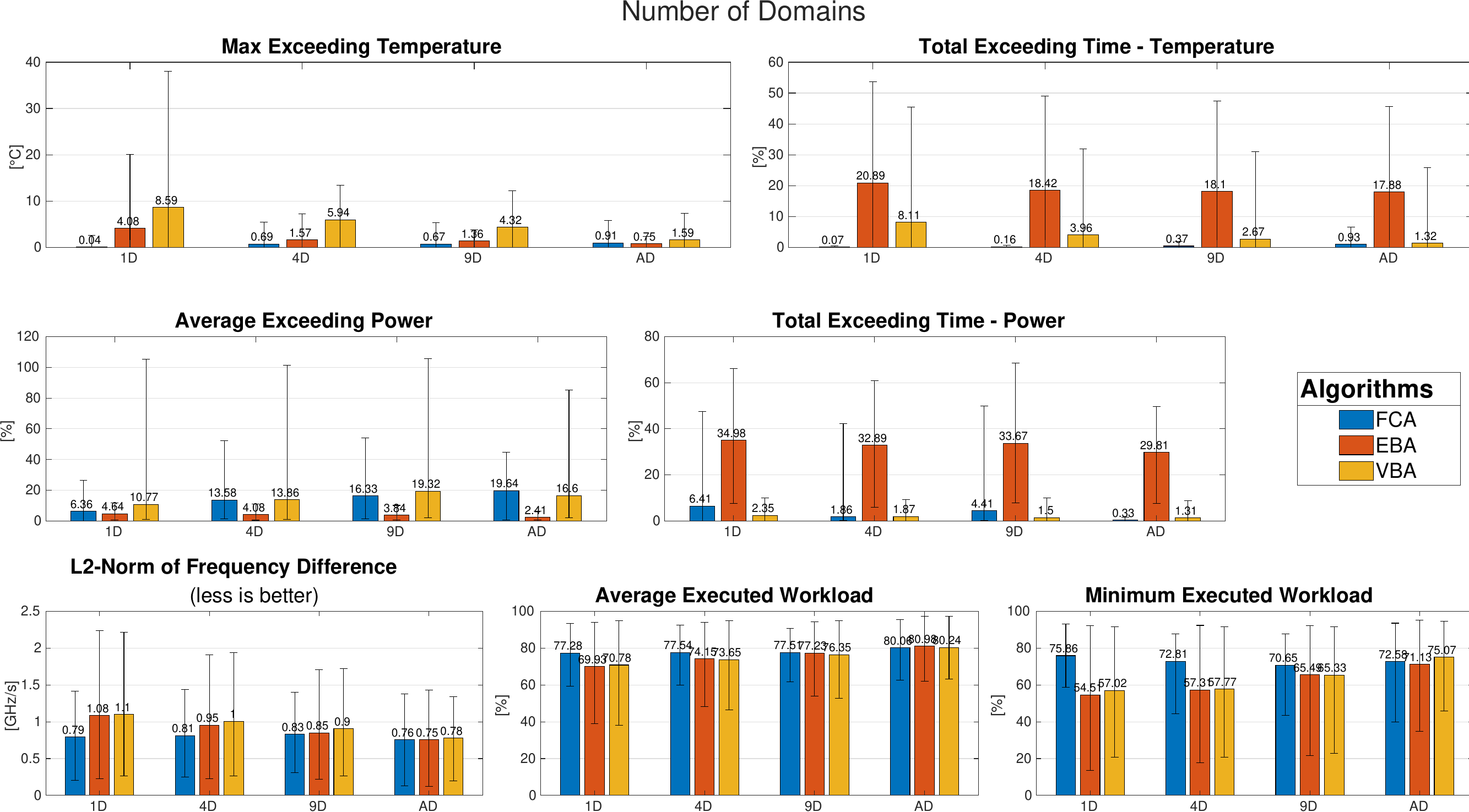}
	\caption{\label{fig:dom} Control performance metrics plotted per number of domains (one -1D-, four -4D-, nine -9D-, and one domain per core -AD-) with their relative error bars. In the first row are the Thermal metrics, in the middle row are the Power metrics, and in the last row are the Target Compliance and execution progress metrics. }
	%\vspace{-15pt}
\end{figure*}

\begin{figure*}[t]
	\centering
	\includegraphics[width=1\linewidth]{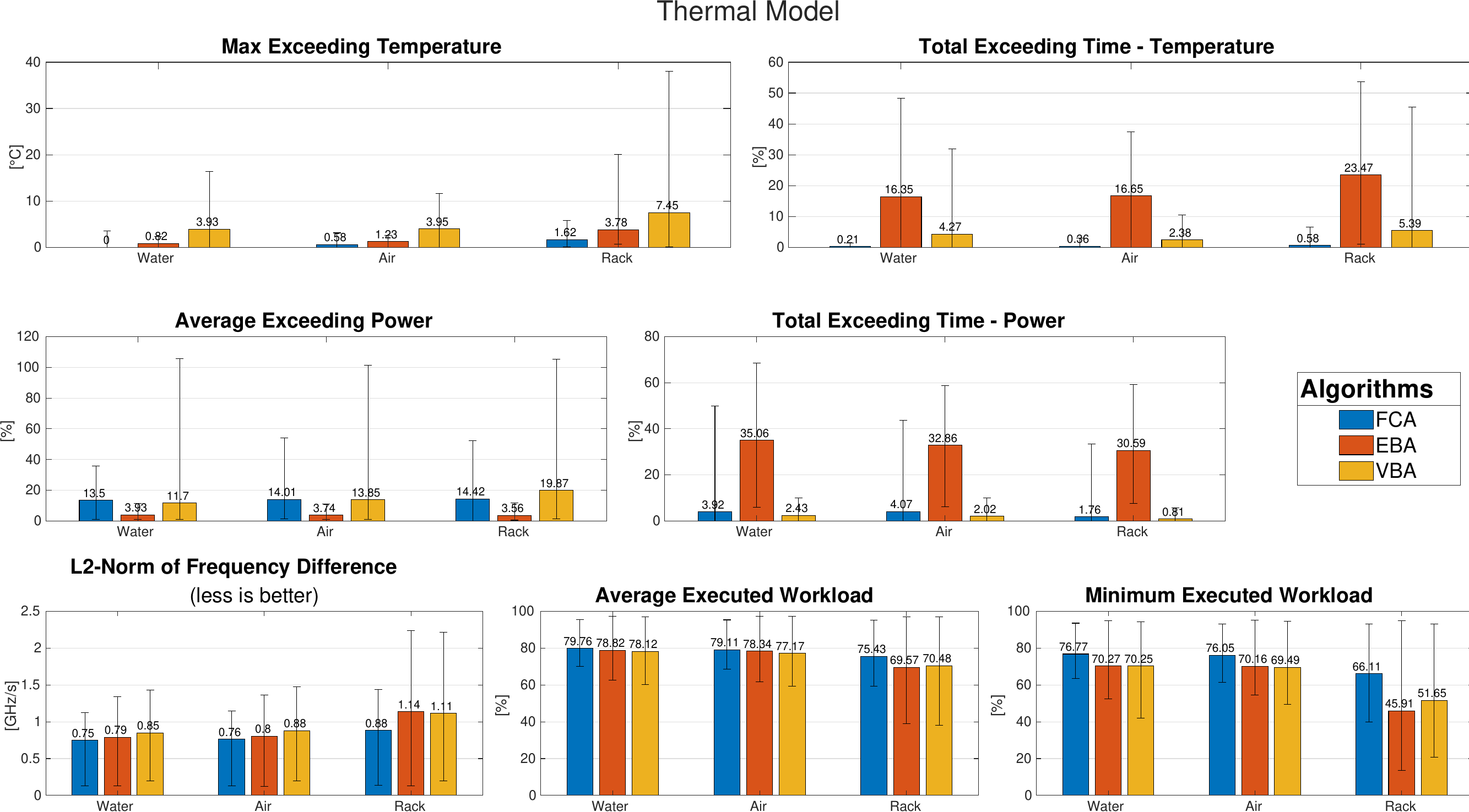}
	\caption{\label{fig:mod} Control performance metrics plotted per Thermal Model type (water cooling -WATER-, air cooling -AIR-, and horizontal rack cooling, -RACK-) with their relative error bars. In the first row are the Thermal metrics, in the middle row are the Power metrics, and in the last row are the Target Compliance and execution progress metrics. }
	%\vspace{-15pt}
\end{figure*}

\begin{figure*}[t]
	\centering
	\includegraphics[width=1\linewidth]{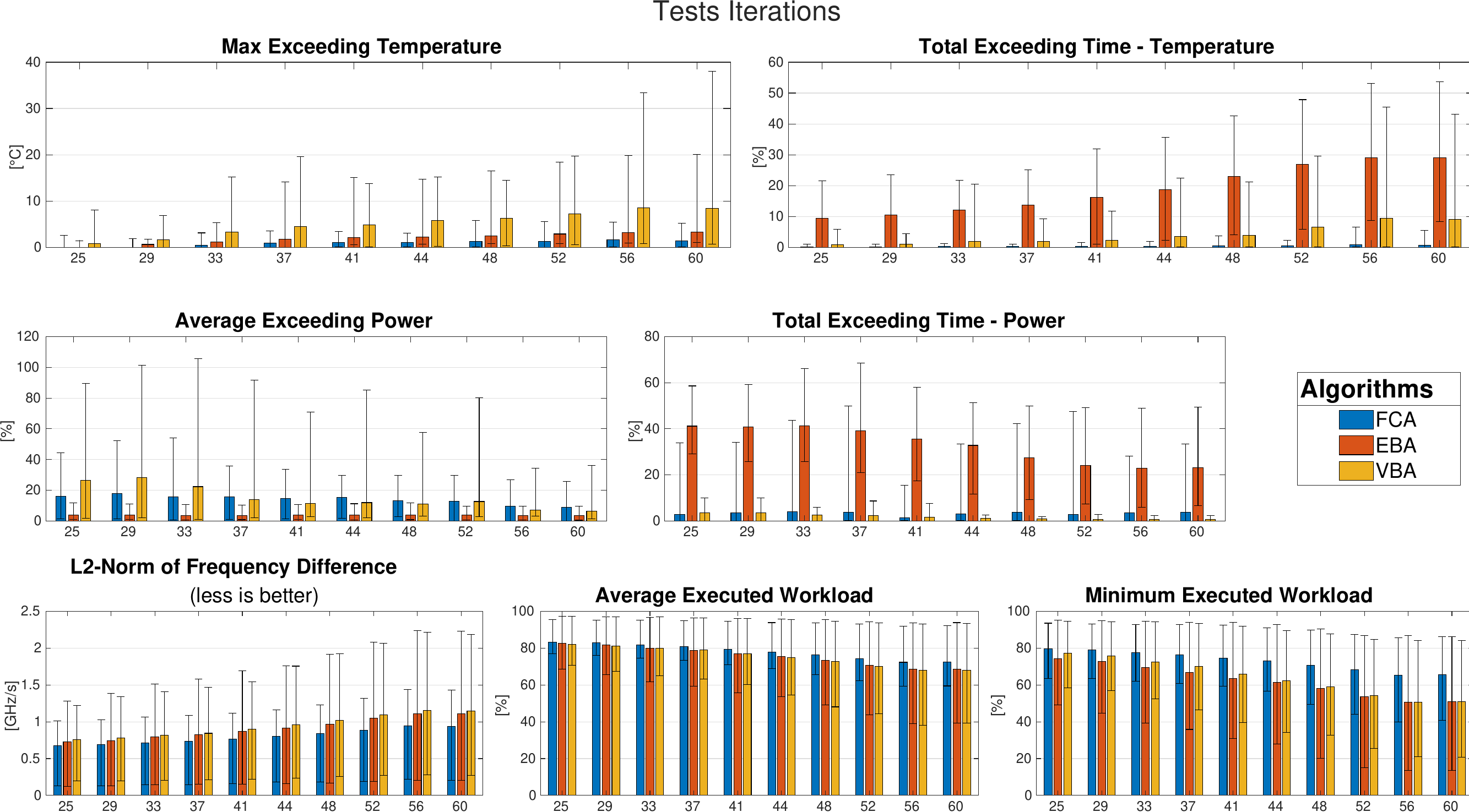}
	\caption{\label{fig:test} Control performance metrics plotted per tests iteration with their relative error bars. In the x-axis are indicated the average initial temperatures of the system. In the first row are the Thermal metrics, in the middle row are the Power metrics, and in the last row are the Target Compliance and execution progress metrics. }
	%\vspace{-15pt}
\end{figure*}

\subsubsection{Power Capping Metrics}
In the power capping metrics, \reb{the \gls{fca} has a high average exceeding power ($13.98\%$), but with a low exceeding time ($3.25\%$ on average). This suggests that the power capping performance degradations are mostly happening on target or application phase changes. The average exciting power distribution has a larger deviation than the other two algorithms and the exceeding time has a large number of outliers which originated from the MULTI-WL tests as \cref{fig:wl} suggests. Unusually, the higher the number of domains, the worse the average exceeding power performance gets, suggesting that the problem may be rooted in oscillations.

\Gls{vba} has a similar behavior as \gls{fca}, with the worst average exceeding power ($15.14\%$) and the lowest exceeding time ($1.76\%$ on average). The average exceeding power has a large number of outliers, reaching up to $105.84\%$ of the power budget, while the exceeding time is the most consistent among the three algorithms. These outliers are originated from the MULTI-WL tests.

\gls{oca} has an opposite behavior to the other two algorithms, with the best and most consistent average exceeding power ($3.74\%$), and the worst exceeding time ($32.84\%$ on average) with the larger deviation.}

%performs worse than both \gls{fca} and \gls{vba}. 
%Even for a small acceptable amount of time (always below $2.18\%$) compatible with a transient settlement region, it exceeds the power budget by maximum values that are often above $50\%$ and averages that are often above $10\%$ of its value. The best performance is achieved with the WATER thermal model.

%\Gls{fca} and \gls{vba} compare favorably in terms of power capping performance. The former has lower exceeded times in MAX-WL and CLOUD-WL (average: $1.09\%$ FCA, $2.88\%$ VBA), but it performs slightly worse in the exceeded values metric (average:$2.39\%$ FCA, $2.01\%$ VBA), mainly in the RACK model tests. In the MULTI-WL test, \gls{vba} outperforms \gls{fca} with almost no exceeded power across all configurations.

%\Gls{fca} has some spurious results (such as all the CLOUD-WL 1-D tests and several A-D tests), meaning that, differently from the thermal capping, the behavior of the power capping part of the algorithm may be more inconsistent. 

\subsubsection{\reb{Target Compliance and Execution Metrics}}
\reb{
On average, the \gls{fca} has the best target compliance across all tests, with the exception of the all-domains (AD) configuration, where the \gls{oca} is only $0.01 GHz/s$ better. In particular, it has a $12.26\%$ lower L2-norm than \gls{oca} and $15.67\%$ lower than \gls{vba}.
The Average application execution progress (AV-$Wl_p$) with the \gls{fca} is $3.34\%$ higher than \gls{oca} and $3.77\%$ higher than \gls{vba}, while the minimum application execution progress (MIN-$Wl_p$) is $17.50\%$ and $14.39\%$ better in \gls{fca} than \gls{oca} and \gls{vba} respectively.

Both the \gls{oca} and \gls{vba} algorithms improve their target compliance and application progress results as the number of domains grows, managing to reach the performance of the \gls{fca} in the AD configuration, confirming the capability of \gls{oca} in absence of voltage coupling. In the MULTI-WL workload test, the \gls{vba} has a better AV-$Wl_p$ result but does worse in target compliance and MIN-$Wl_p$. This is because} \gls{vba} tends to homogenize frequencies among cores, independently of the workload, favoring the less power-hungry (in this case the cores running the mix of int and floating point workload at $2.7GHz$). Instead, \gls{fca} and \gls{oca} algorithms favor the cores running the more demanding workload, which, at the same conditions ($F$, $V$, $T$), consume more power and generates more heat than other workloads~\cite{bartolini2019advances, schone_energy_2019}. %1.39\% and 1.92\% respectively

\reb{
Regarding results variation, \Cref{fig:violinplot} shows that \gls{fca} has the smaller deviation in the average and minimum execution progress, while all three algorithms have a bimodal target compliance distribution. This is caused by the MULTI-WL tests. These tests have the majority of cores in a low-power state, thus leaving thermal and power room for the other cores to execute faster, and generating a set of tests in which all three algorithms perform better than the other types of workload (MAX-WL and CLOUD-WL). Figures (\ref{fig:dom}, \ref{fig:mod}, \ref{fig:test}) also shows that the \gls{fca} is the most consistent regarding target compliance and execution progress across different domains, models, and initial conditions respectively, with the variation being mostly regarding different workload. This is confirmed by \cref{fig:wl} where the error bars of \gls{fca} are significantly smaller than the other algorithms.
}
%On a final note, all algorithms have significantly increased performance in the A-D tests, meaning that the distribution of the two groups of cores in the different domains was able to stress the decision capability of each control algorithm with domain voltage coupling.

\subsubsection{\reb{Summary}}

The average results and their standard deviation to account for results' consistency are summarized in \cref{tbl:final_res}.

\definecolor{Green}{rgb}{0,0.901,0}
\definecolor{Red}{rgb}{0.949,0,0}

%%%%%%%%%%%%%%%%%%%%%%%%%%%%%%%%%
% https://www.latex-tables.com/
%%%%%%%%%%%%%%%%%%%%%%%%%%%%%%%%%

% \usepackage{color}
% \usepackage{rotating}
% \usepackage{tabularray}
% \usepackage{color}
% \usepackage{rotating}
% \usepackage{tabularray}
\definecolor{Green}{rgb}{0,0.901,0}
\definecolor{Red}{rgb}{0.949,0,0}
\begin{table*}
\centering
\label{tbl:final_res}
\begin{tblr}{
  row{odd} = {c},
  row{2} = {c},
  row{4} = {c},
  row{6} = {c},
  row{8} = {c},
  row{10} = {c},
  row{14} = {c},
  row{16} = {c},
  cell{1}{1} = {c=3}{},
  cell{2}{1} = {r=6}{},
  cell{2}{2} = {r=3}{},
  cell{2}{4} = {Green},
  cell{2}{6} = {Red},
  cell{3}{4} = {Green},
  cell{3}{6} = {Red},
  cell{4}{4} = {Green},
  cell{4}{6} = {Red},
  cell{5}{2} = {r=2}{},
  cell{5}{4} = {Green},
  cell{5}{5} = {Red},
  cell{6}{4} = {Green},
  cell{6}{5} = {Red},
  cell{7}{2} = {c=2}{},
  cell{8}{1} = {r=4}{},
  cell{8}{2} = {r=2}{},
  cell{8}{5} = {Green},
  cell{8}{6} = {Red},
  cell{9}{5} = {Green},
  cell{9}{6} = {Red},
  cell{10}{2} = {r=2}{},
  cell{10}{5} = {Red},
  cell{10}{6} = {Green},
  cell{11}{5} = {Red},
  cell{11}{6} = {Green},
  cell{12}{1} = {r=6}{},
  cell{12}{2} = {r=2}{c},
  cell{12}{3} = {c},
  cell{12}{4} = {Green,c},
  cell{12}{5} = {c},
  cell{12}{6} = {Red,c},
  cell{13}{4} = {Green},
  cell{13}{5} = {Red},
  cell{13}{6} = {Red},
  cell{14}{2} = {r=2}{},
  cell{14}{4} = {Green},
  cell{14}{6} = {Red},
  cell{15}{4} = {Green},
  cell{15}{6} = {Red},
  cell{16}{2} = {r=2}{},
  cell{16}{4} = {Green},
  cell{16}{5} = {Red},
  cell{17}{4} = {Green},
  cell{17}{5} = {Red},
  vlines,
  hline{1-2,8,12,18} = {-}{},
  hline{3-4,6,9,11,13,15,17} = {3-6}{},
  hline{5,7,10,14,16} = {2-6}{},
}
Alg                                                &                                       &               & \textbf{FCA} & \textbf{EBA} & \textbf{VBA} \\
\hline
\begin{sideways}\textbf{Temperature}\end{sideways} & \textbf{Exc. Max Value [$\degree C$]} & \textbf{Max~} & 5.83         & 20.02        & 38.01        \\
                                                   &                                       & \textbf{Av}   & 0.58         & 1.94         & 5.11         \\
                                                   &                                       & \textbf{SD}   & 2.83         & 3.36         & 5.28         \\
                                                   & \textbf{Exceeded Time [$\%$]}         & \textbf{Av}   & 0.38         & 18.82        & 4.01         \\
                                                   &                                       & \textbf{SD}   & 0.73         & 12.10        & 7.09         \\
                                                   & \textbf{ Average Temp [$\degree C$]}  &               & 67.87        & 71.96        & 69.23        \\
\hline
\begin{sideways}\textbf{Power}\end{sideways}       & \textbf{Exc. Av Value [$\%$]}         & \textbf{Av~}  & 13.98        & 3.74         & 15.14        \\
                                                   &                                       & \textbf{SD}   & 10.65        & 3.45         & 25.16        \\
                                                   & \textbf{Exceeded Time [$\%$]}         & \textbf{Av}   & 3.25         & 32.84        & 1.76         \\
                                                   &                                       & \textbf{SD}   & 8.22         & 13.27        & 2.11         \\
\hline
\begin{sideways}\textbf{Execution}\end{sideways}   & \textbf{2-norm [$GHz/s$] }            & \textbf{Av}   & 0.80         & 0.91         & 0.95         \\
                                                   &                                       & \textbf{SD}   & 0.35         & 0.45         & 0.45         \\
                                                   & \textbf{AV-$Wl_p$~[$\%$] }            & \textbf{Av}   & 78.10        & 75.57        & 75.26        \\
                                                   &                                       & \textbf{SD}   & 8.55         & 12.01        & 12.58        \\
                                                   & \textbf{\textbf{MIN-$Wl_p$~[$\%$]}}   & \textbf{Av}   & 72.98        & 62.11        & 63.80        \\
                                                   &                                       & \textbf{SD}   & 10.86        & 18.68        & 17.16        
\end{tblr}

\caption{Summary of the average results of the battery of tests. Standard deviation is included to show the constancy of each algorithm among different tests, models, and configurations. In green the best result, in red the worst. }
\end{table*}

\reb{The \gls{fca} achieves the best target compliance and application progress while outperforming the other algorithms in thermal capping and being able to cap power comparably to the \gls{vba}.
This means that not only the \gls{fca} applies the given \gls{hlc} target better, but it also makes the applications run around $3\%$ faster and with lesser thermal room. Additionally, the execution progress across \glspl{pe} is more consistent, as the application with the minimum progress is closer to the average value with respect to the other algorithms. The \gls{fca} has also the most consistent behavior across different cooling configurations, number of domains, and initial conditions, as the error bars in \cref{fig:wl} clearly show.

\gls{oca} total exceeding time distribution spread confirms that this algorithm doesn't have a consistent behavior when non-idealities are introduced into the model, as found also in \cref{fig:old_results}. 
Specifically, \gls{oca} and \gls{vba} encounter significant challenges with larger domains, highlighting the difficulties in managing voltage coupling. Conversely, the \gls{fca}’s iterative root-finding method demonstrates better precision in achieving the desired operating point.

Power capping results suggest that it is difficult to effectively cap power and there is a trade-off between exceeding time and exceeding value, as all three algorithms had opposite behavior on those metrics
}. 

%The worst thermal capping performance in the cloud-wl indicate the difficulty to manage varying and unpredictable workload

%multi-wl tests created the worst power capping performnace across all algorithm suggesting domains divided as ...

%

\subsection{\Gls{hil} evaluation}

While the focus of this work is to provide a cross-evaluation between several algorithms and thermal models of a modern computing \gls{hpc} processor (\cref{fig:chip}), we complete the assessment by verifying the feasibility of the proposed algorithms on the \gls{hil} framework proposed in~\cite{Ottaviano2023ControlPULPAR} to exploit real hardware interfaces.

\reb{On the controller side, the adoption of an \gls{fpga} emulation hosting the actual power controller \gls{hw} offers significant advantages, such as the ability to test control firmware developed during the software- and model-in-the-loop (\gls{sil} and \gls{mil}, respectively) design phases on actual hardware controllers with the assurance of cycle-accurate simulation. This precision is essential for achieving detailed hardware observability and controllability~\cite{HIL_ISUT}, as well as ensuring a one-to-one correspondence between the \gls{rtl} source and its \gls{fpga} implementation in terms of clock cycles, exemplified by direct FAME systems~\cite{FAME}.}

The framework consists of a Xilinx Zynq Ultrascale+ \gls{fpga}-\gls{soc} that hosts both the thermal and power model for the system under control and the \gls{llc}. We run the \gls{fca} on top of the \gls{llc} with a hyperperiod of $500us$. 
%While being able to deploy and execute \gls{oca} in~\cite{Ottaviano2023ControlPULPAR}, the newly proposed \gls{fca} proves to have an optimal memory footprint that is compatible with resource-constrained embedded systems.
%The small amount of LUT demanded by the algorithm incurs a negligible overhead compared to the available memory budget $\le1MiB$ of \gls{sram}.

\reb{
On a single core running at $50MHz$, a basic version of the \gls{llc} firmware controls a chiplet of 24 cores, occupying up to the $78.18\%$ of the hyperperiod. The \gls{fca} algorithm occupies on average the $96.3\%$ of those cycles. In particular, of those cycles, the initialization contributes to the $9.57\%$, the Fuzzy-inspired thermal capping to the $10.54\%$, the Conv2P and the power capping to the $29.09\%$, and the iterative Conv2F part to the $48.79\%$. The remaining $2.02\%$ comes from the sum of other initialization and management parts.}

\reb{From this data, we can observe how the control algorithm is costly, and that the parts relative to the power conversion (Conv2P and Conv2F) take more than $50\%$ of the whole algorithm. For this reason, the \gls{llc} \gls{hw} proposed in~\cite{Ottaviano2023ControlPULPAR} includes a cluster accelerator} comprising 8 32-bit RISC-V cores to accelerate the more demanding fragments of the algorithm, such as the root-finding bisection method.

\reb{The memory footprint of this basic firmware version is $120KiB$. This includes initialized data, instructions, and uninitialized data. We observe that this fits the memory requirements of \glspl{llc}, typically below 1MiB.

This experiment demonstrates that (i) \gls{hw} constraints such as execution speed and memory footprint are crucial to fully validate the developed controller algorithm; (ii) increasingly complex algorithms can be successfully implemented in \gls{hw}, but timing constraints (control period) should be adjusted accordingly to fit the computation; (iii) more computationally-demanding algorithms, such as model-based or \gls{ml}-driven, would be challenging to implement on actual \glspl{llc}, hence demanding for the integration of general-purpose or domain-specific \gls{hw} acceleration techniques to meet the imposed constraints~\cite{Ottaviano2023ControlPULPAR}.}

\section{Conclusion}\label{sec:conclusions}

This paper presents the modeling of an \gls{hpc} chip that takes into account the non-linearities and non-idealities of a real system and analyzes the control problem and challenges that arise from the realistic scenario.
We first extend the work presented in Bambini et al.~\cite{CCTA_bambini} to satisfy the introduced constraints and coupling in the thermal and power model. %
Then, we propose a novel control design inspired by the fuzzy control theory for thermal capping. %
Finally, we implement an iterative solving method to solve the coupling constraint instead of moving average filters.

The presented control design outperforms the \gls{pid} algorithms in thermal capping by reducing the average exceeded temperature by up to $5\times$ while being able to deliver \reb{more than $3\%$} better application execution performance. Across all tests and configurations executed within a \gls{mil} framework, it proves to be the algorithm achieving the most stable results. 
The larger performance gains compared to the other algorithms happen in the scenario with larger domains (1-D and 4-D) and the RACK model, where the thermal coupling is more complex. %
The analysis of the results suggests that the proposed design is beneficial in addressing the control challenges presented in \cref{ssec:ctrl_challenges}. The original design from~\cite{CCTA_bambini} is still competitive in the A-D configuration, being the more compliant with the assumptions made in the original work.

While the fuzzy-inspired design solves the thermal capping problem, the power distribution block is still overly complex: it is model-based, it assumes to have a measure of the workload $\omega$, and it does not fully take into account the output coupling. Future works in this direction should investigate a simpler and more comprehensive power distribution block. Model predictive solutions could be considered, provided their efficient deployment on resource-constrained \glspl{llc}.

%in the future we could investigate a different solution, maybe a research tree algorithm with a lut with all the possible FV combination, that takes into account power distribution, fv releation, and shared V. Still this solution could be too computationally expensive and alsoo too monholitic to parallelize proficiently.

%\printbibliography

\section{Acknowledgments}
\ifx\blind\undefined
    The study has been conducted in the context of 
    EU H2020-JTI-EuroHPC-2019-1 project REGALE (g.n. 956560), 
    EuroHPC EU PILOT project (g.a. 101034126), 
    EU Pilot for exascale EuroHPC EUPEX (g. a. 101033975), and 
    European Processor Initiative (EPI) SGA2 (g.a. 101036168).
\else
    \textit{Acknowledgments omitted for blind review.}
\fi

%%%%%%%%%%%%%%%%%%%%%%%
%%%   BACK MATTER   %%%
%%%%%%%%%%%%%%%%%%%%%%%

%\printglossary[type=\acronymtype]

\bibliographystyle{IEEEtran}
\bibliography{bibliography.bib}

\end{document}